\documentclass[10pt]{article}
\usepackage{times}
\RequirePackage{natbib}
\usepackage[colorlinks=true, citecolor=blue, linkcolor=blue]{hyperref}
\usepackage{amsmath, amssymb, fullpage, amsthm, array,,graphicx,asa, url}

\graphicspath{{images/}}

\usepackage{color}
\definecolor{mahbub}{rgb}{1.0, 0.6, 0.4}
\newcommand{\hh}[1]{{\color{magenta} #1}}
\newcommand{\dc}[1]{{\color{green} #1}}
\newcommand{\mm}[1]{{\color{mahbub} #1}}

\evensidemargin 
\oddsidemargin

\begin{document}


\title{Human Factors Influencing Visual Statistical Inference }
\author{{Mahbubul Majumder, Heike Hofmann, Dianne Cook}
\thanks{Mahbubul Majumder is an Assistant Professor in the Department of Mathematics, University of Nebraska at Omaha, NE 68182 (e-mail: mmajumder@unomaha.edu), Heike Hofmann and Dianne Cook are Professors in the Department of Statistics and Statistical Laboratory, Iowa State University, Ames, IA 50011-1210. This research is supported in part by the National Science Foundation Grant \# DMS 1007697.}}
\date{\vspace{-.5in}}
\maketitle

\begin {abstract}  
Visual statistical inference is a way to determine significance of patterns found while exploring data. It is dependent on the evaluation of a lineup, of a data plot among a sample of null plots, by human observers. Each individual is different in their cognitive psychology and judiciousness, which can affect the visual inference. The usual way to estimate the effectiveness of a statistical test is its power. The estimate of power of a lineup can be controlled by combining evaluations from multiple observers. Factors that may also affect the power of visual inference are the observers' demographics, visual skills, and experience, the sample of null plots taken from the null distribution, the position of the data plot in the lineup, and the signal strength in the data. This paper examines these factors. Results from multiple visual inference studies using Amazon's Mechanical Turk are examined to provide an assessment of these. The experiments suggest that individual skills vary substantially,  but demographics do not have a huge effect on performance. There is evidence that a learning effect exists but only in that observers get faster with repeated evaluations, but not more often correct. The placement of data plot in the lineup does not affect the inference.

\end {abstract}


\section{Introduction}  

The lineup protocol introduced in \citet{buja:2009} can be used to test the significance of findings during the exploratory data analysis. The methodology is a part of what is called visual statistical inference.  These concepts have been developed further by \citet{majumder:2013} who refined the terminology and validated the lineup protocol with a head to head comparison with conventional inference. One of the major contributions of  \citet{majumder:2013}  is to define the power of the visual test and provide methods to obtain the power for a particular lineup. It was observed that the power can be as good or better than that of a conventional test in some scenarios.

In visual inference, the test statistic is a plot of the observed data. To create a lineup, this plot, called the actual data plot, is placed in a layout of null plots. The null plots  are generated from the model specified by a null hypothesis, essentially describing what the plot might look like if the data had no structure. An observer is asked to evaluate the lineup. If the  actual data plot is detected by the observer, the null hypothesis is rejected. This means that the structure in the actual data plot has significant structure, a pattern that is not simply due to randomness. Combining the choices of multiple observers provides more stability in the estimation of significance.


\begin{figure}[hbtp] 
   \centering
   \includegraphics[width=0.99\textwidth]{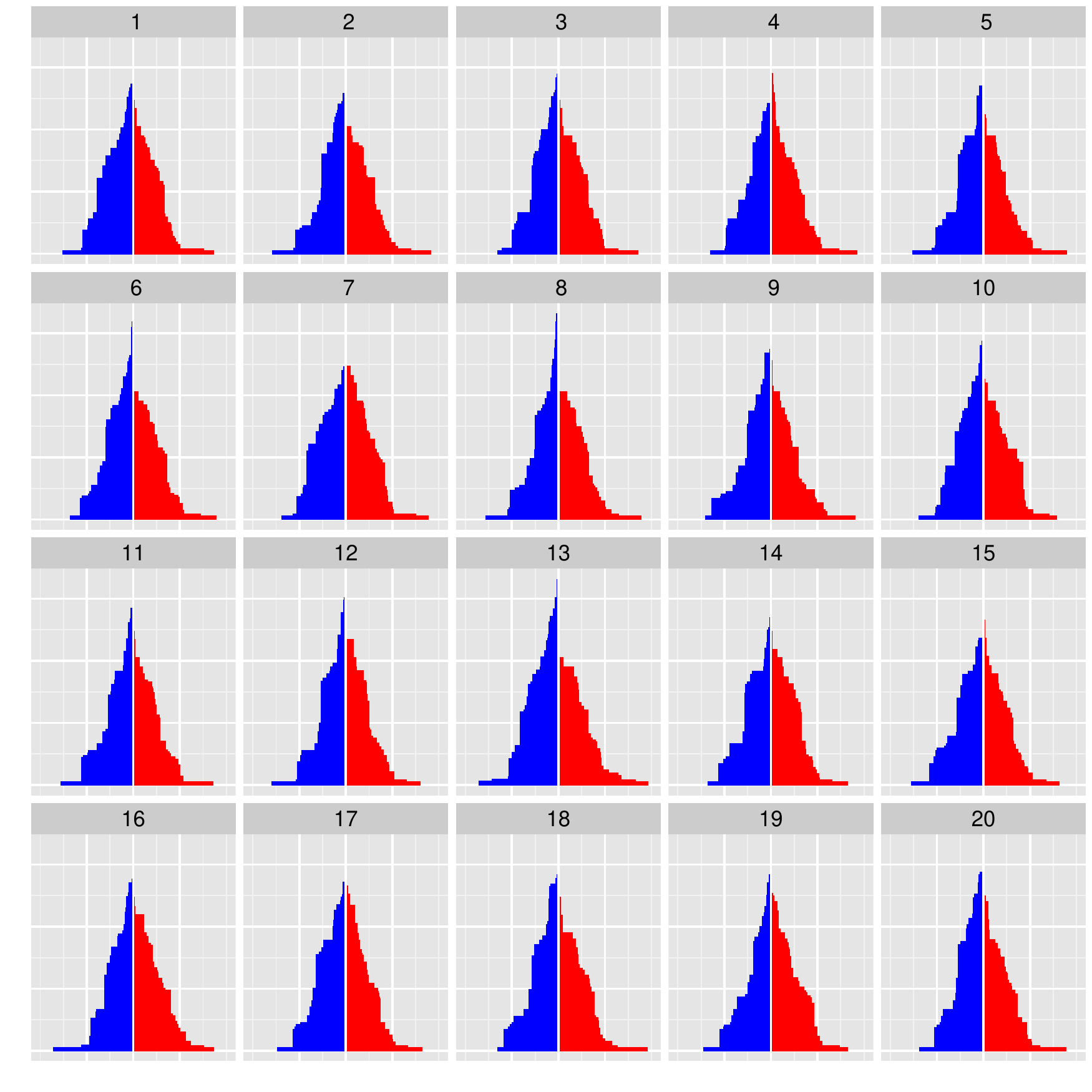} 
   \caption{ \label{fig:elect-1} Which one of the plots is the most different from the others?}
\end{figure}

\begin{figure}[hbtp] 
   \centering
   \begin{minipage}[c]{0.49\textwidth}
   \includegraphics[width=\textwidth]{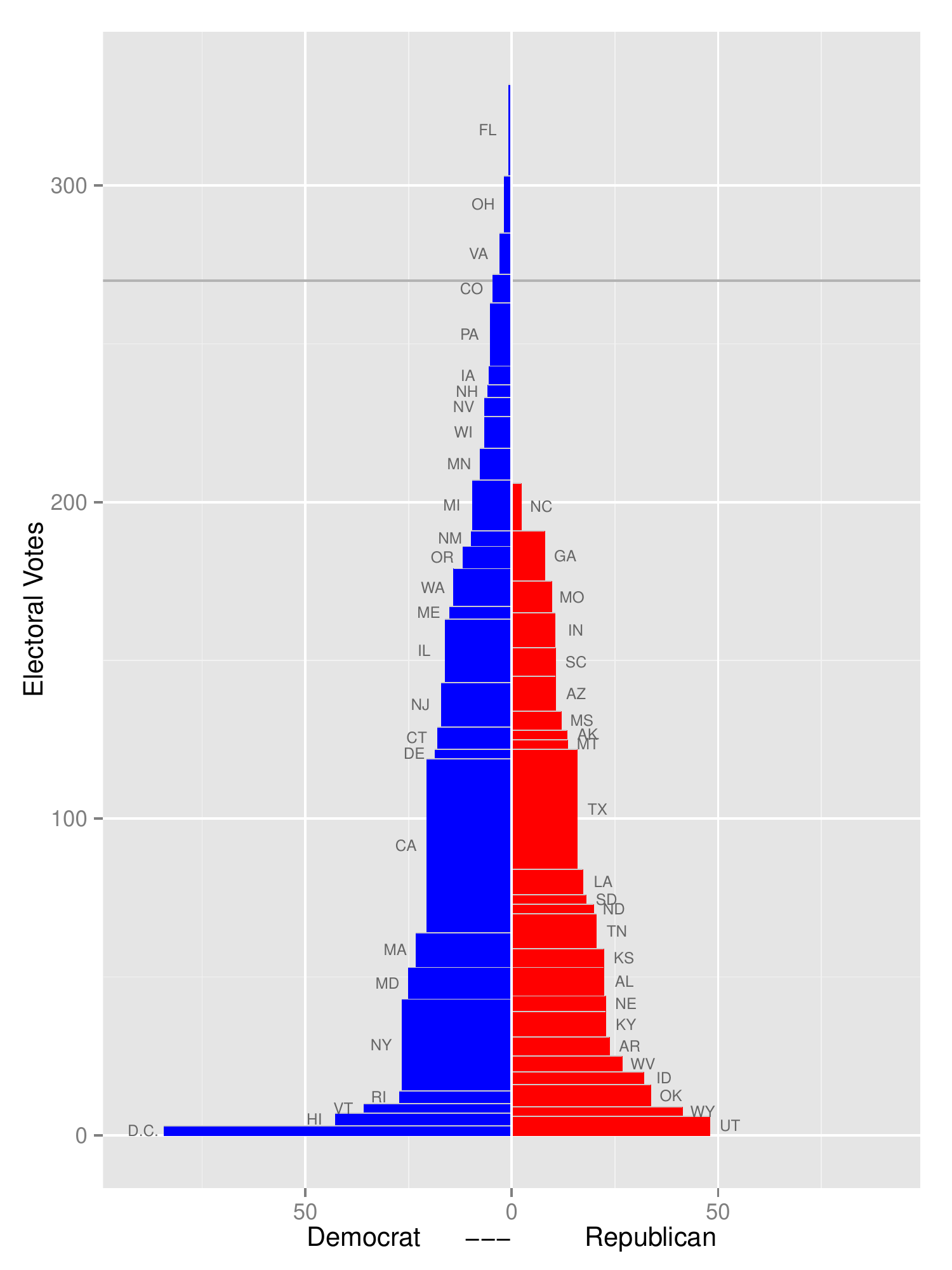}
   \end{minipage}
    \hfill
   \begin{minipage}[c]{0.45\textwidth}
   Poll aggregator 1 \\
      \includegraphics[width=\textwidth]{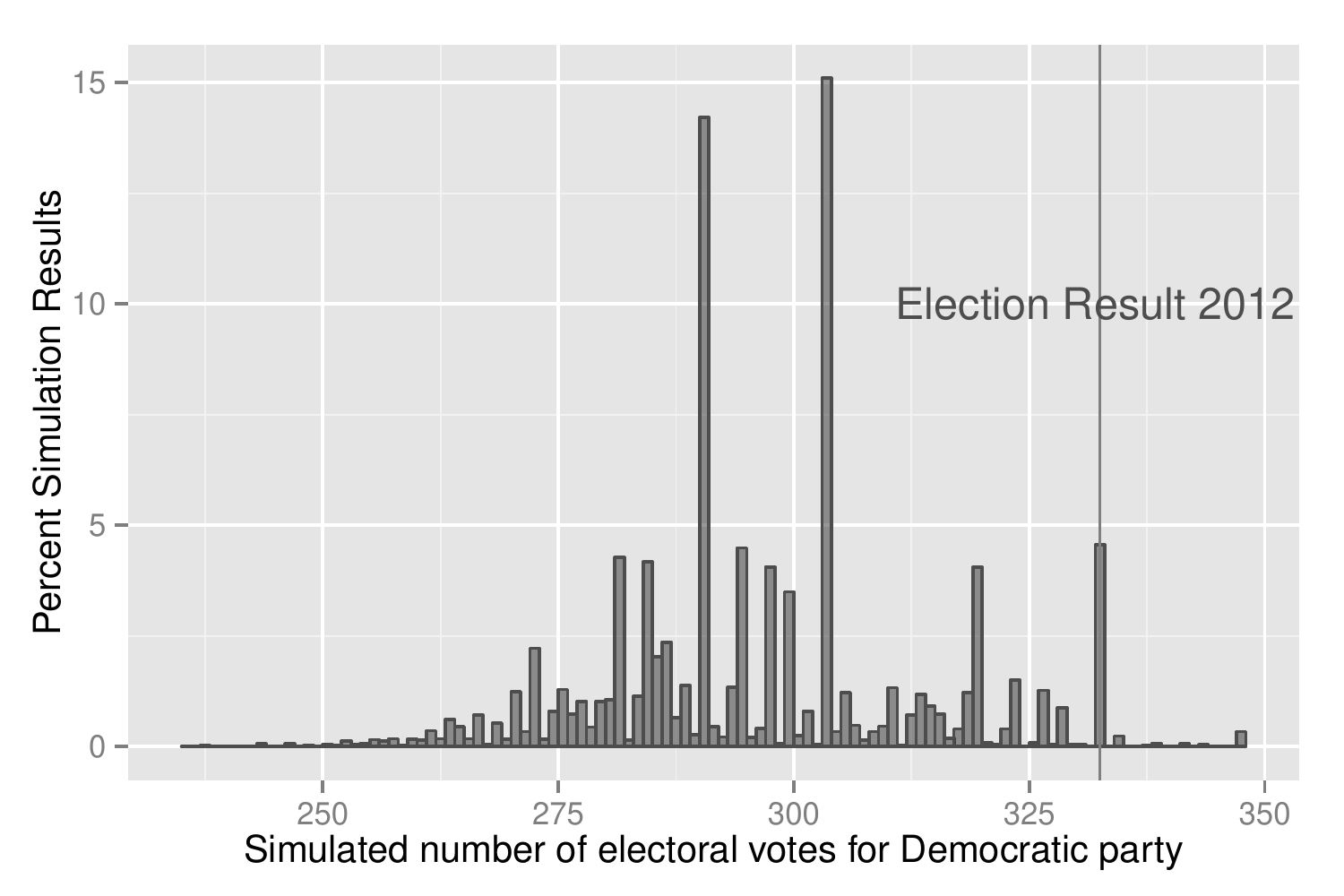} 
   Poll aggregator 2\\
      \includegraphics[width=\textwidth]{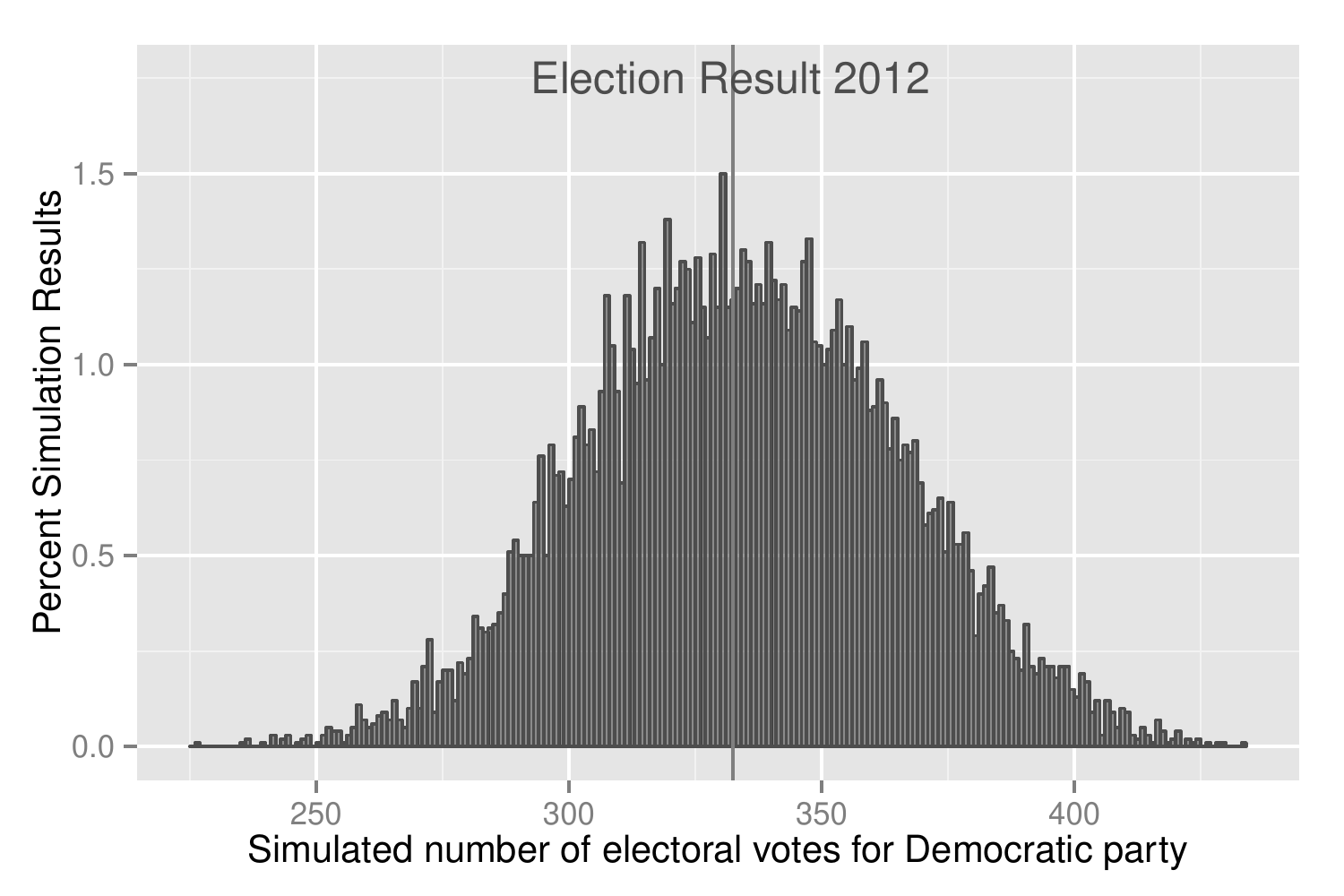} 
   \end{minipage}
   \caption{ \label{fig:tower} Electoral building plot of the results of the 2012 U.S. Presidential Election (left). On the right two histograms of 10,000 simulations each based on polling averages from two different sources. For the histogram on the top, the $p$-value of observing results as extreme as  the  2012 U.S. election results based on the bootstrap is 0.0533 (with Bootstrap standard error of 0.002), making the election results almost significantly different from the polls (data source for poll aggregator 1 is Freedom`s Lighthouse Averages). There is no indication of any inconsistency between polls and election results based on the bootstrap simulation below (data source for poll aggregator 2 is RealClearPolitics Averages). The lineups are based on the top source.}
\end{figure}

Figure~\ref{fig:tower} displays a lineup of 20 plots where one of the plots is observed data, while the remaining 19 plots are showing from data generated under a null model. Which one of the 20 plots is the most different from the others? 
When asked this question, 12 \hh{out} of 72 observers picked the  plot \hh{of the observed data} (located in spot number $3^2+4$ in the lineup).  The corresponding $p$-value is \hh{0.0077}, indicating sufficient evidence to reject the null hypothesis. 

\hh{How do we interpret this finding, though?} For that, we need to know the context of the data and we need to have more information about the \hh{how null plots were generated}. 
This particular example investigates the results from the 2012 US presidential election in comparison to the poll results just prior to the election. 
(Although this example is more simplistic than most of the tests conducted to date, it will serve the purpose of illustrating the lineup protocol.) The data is looking at the difference in poll results between the two (major) presidential candidates, Obama and Romney, for all states. Each panel in Figure~\ref{fig:elect-1} shows an `electoral building' where each state in the union is represented by a rectangle. This difference is plotted horizontally, and the height of each box corresponds to the state's electoral votes. Color indicates party affiliation.  The null hypothesis is ``that the election results were consistent with the polls''. The polling results provides the null model from which data is simulated. Because each poll has a margin of error, this is used to simulate different scenarios that might have resulted on election day, if the polls were on target. A null data set is generated as a set of draws from normal distribution, with mean equal to the difference in poll percentage of the latest state poll results, and standard deviation equal to 2.5, approximating a margin of error of 5\%. These samples are plotted as electoral buildings, and the plot made with the results from the election is placed randomly among them in a lineup of size $5\times 4$. If the null hypothesis is true the actual data plot should look just like any of the other plots, and not be identified by an observer. Figure~\ref{fig:tower} shows a plot of the electoral building with added context information and labels.

A lineup can be evaluated by a single person or multiple observers. A distribution \hh{similar to the binomial distribution, but adjusted for dependencies introduced by the lineup scenario,} is used to calculate the $p$-value based on the number of times observers identify the actual data plot, which provides the information needed to make a decision on rejecting or failing to reject the null hypothesis.  In order to avoid expectation errors \citep{meilgaard}, observers should not be aware of the data that constitutes a lineup, and should not have seen the actual data plot before seeing the lineup. This is the reason that in the election example, above, the scenario was explained after the lineup question, in the text. 

The question that is asked of the observer should be as general as possible, effectively asking the observer to pick the plot that is different, and allowing them to provide their reasons for seeing their pick as different. In some of the studies, the ones described in  \cite{majumder:2013} very specific questions were asked, because the experiments were being conducted to compare results from the lineup protocol with those of conventional tests. In those experiments, structure in the data was strictly controlled in the simulation process, which allowed for specific questions to be asked. In the election example, observers were asked ``which plot is the most different?''. The type of plot, showing two (modified) stacked bar charts in different colors should suggest to the observer that the interesting feature is most difference between the two heights. Most observers got this, but it is possible that some observers might pick plot 4, where the red tower is slightly above the blue as the most different because it is the only plot with this feature. So a better question may have been ``Which plot shows the biggest height difference between the two towers?'' except that this tailors the inference to a specific feature which does not match the null hypothesis of interest. 



For any evaluation the observer may or may not identify the actual data plot. Under the null hypothesis, the actual data plot should look similar to the null plots making it harder to detect. It is not expected that an observer would be able to detect the data plot in this scenario. But since there are limited number of plots in a lineup, which is 20 in the election example, there is a 1/20 chance that the observer would pick the actual plot. This proportion is associated with the Type~I error of the test. On the other hand, if null hypothesis is not true, the observed plot should look different from the null plots, making it easier to be detected. This is the definition of the power of the test. When multiple observers evaluate a lineup, the detection rate can be used to estimate the power. The ability of individual observers can vary, and examining this is the purpose of this paper.




\begin{table*}[hbtp] 
\centering 
\caption{Overview of 10 different Turk experiments, from where data was taken to study human factor effects. All of the experimental data was used to estimate the effect of demographic factors (DF) on visual inference while three were suitable for assessing learning trend (LT) and location effect (LE) was possible to assess using just one specially designed study.} 
\begin{tabular}{m{.5cm}m{2.6cm}m{2cm}m{5.5cm}m{3cm}} 
\hline\hline 
 & Experiment &  Test Statistic  & Lineup question & Used in study of\\ [0.5ex] 
\hline 
1  & Box plot & \begin{minipage}[t]{2cm} \begin{center}	\scalebox{0.12}{\includegraphics{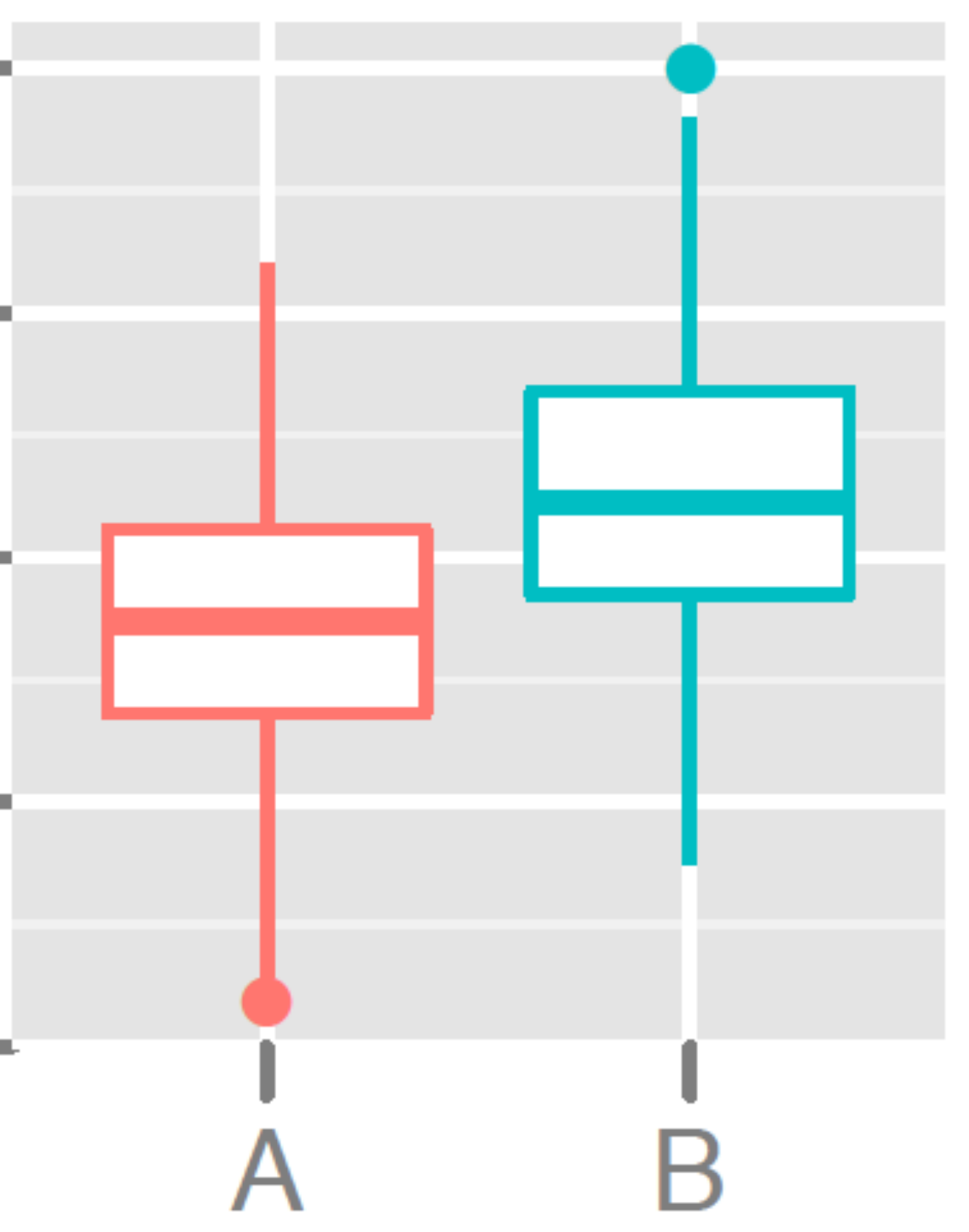}} \end{center} \end{minipage} & Which set of box plots shows biggest vertical difference 
between group A and B? & DF\\
2 &  Scatter plot & \begin{minipage}[t]{2cm}  \begin{center} \scalebox{0.3}{\includegraphics{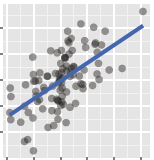}} \end{center} \end{minipage} & Of the scatter plots below which one shows data that has steepest slope? & DF\\
  3 & Contaminated plot &\begin{minipage}[t]{2cm} \begin{center} \scalebox{0.5}{\includegraphics{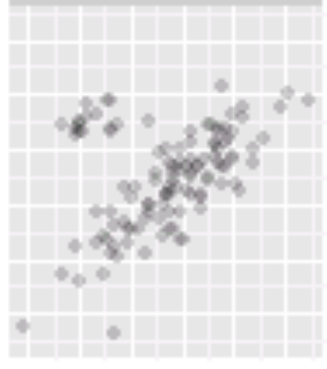}} \end{center} \end{minipage} & Of the scatter plots below which one shows data that has steepest slope? & DF\\
 4 & Polar vs Cartesian & \begin{minipage}[t]{2cm} \begin{center}  \scalebox{0.32}{\includegraphics{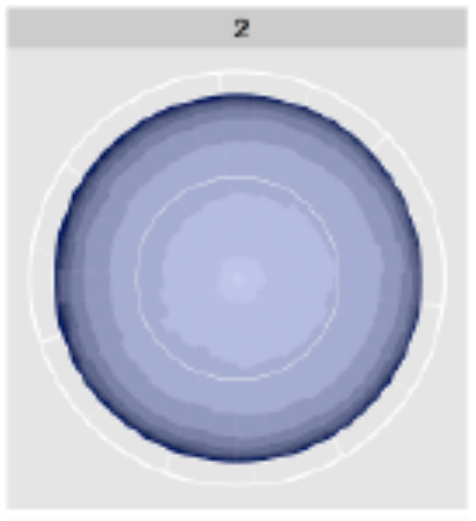}} \end{center} \end{minipage} &Which plot is different?& DF\\
  5 & Hist vs density & \begin{minipage}[t]{2cm} \begin{center}  \scalebox{0.38}{\includegraphics{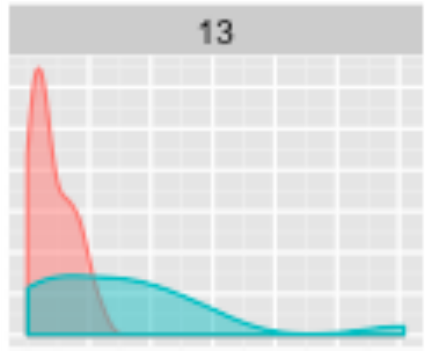}} \end{center} \end{minipage} &In which plot is the blue group furthest to the right?&  DF and LT \\  
  6 & Violin vs boxplot & \begin{minipage}[t]{2cm} \begin{center}  \scalebox{0.35}{\includegraphics{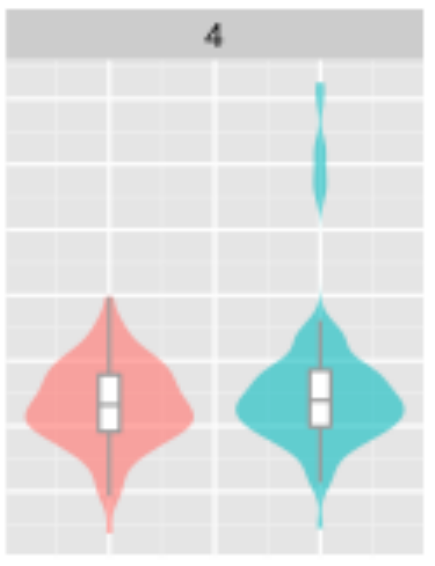}} \end{center} \end{minipage}&  In which plot does the blue group look the most different from  the red group? &  DF and LT \\
  7 & Group separation & \begin{minipage}[t]{2cm} \begin{center}  \scalebox{0.4}{\includegraphics{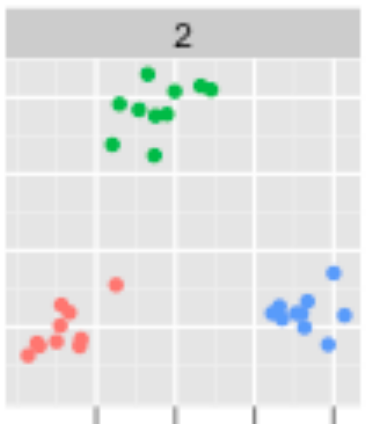}} \end{center} \end{minipage} &Which of these plots has the most separation between the coloured groups? & DF and LT \\ 
  8 & Sine Illusion & \begin{minipage}[t]{2cm} \begin{center}  \scalebox{0.28}{\includegraphics{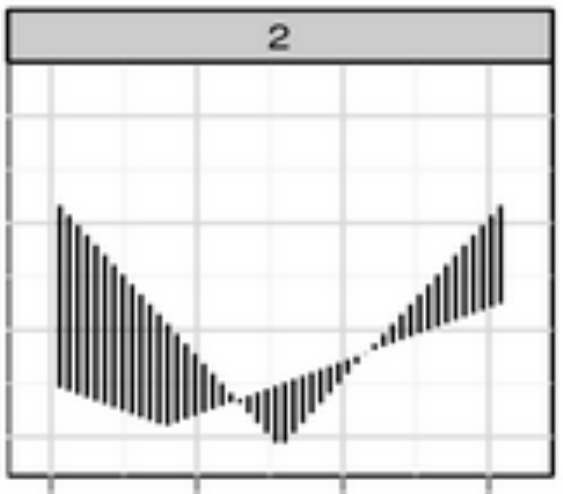}} \end{center} \end{minipage} & In what picture is the size of the curve most consistent? & DF\\ 
  9 & Gene expression &\begin{minipage}[t]{2cm} \begin{center}  \scalebox{0.45}{\includegraphics{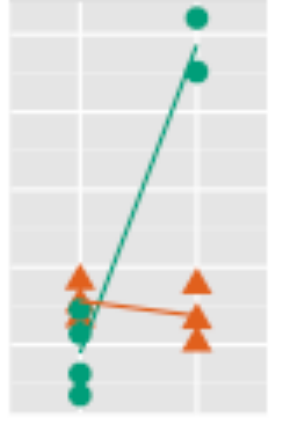}} \end{center} \end{minipage} & In which of these plots is the green line the steepest, and the spread of the green points relatively small? & DF and LE\\ 
  10 & Test normality & \begin{minipage}[t]{2cm} \begin{center}  \scalebox{0.35}{\includegraphics{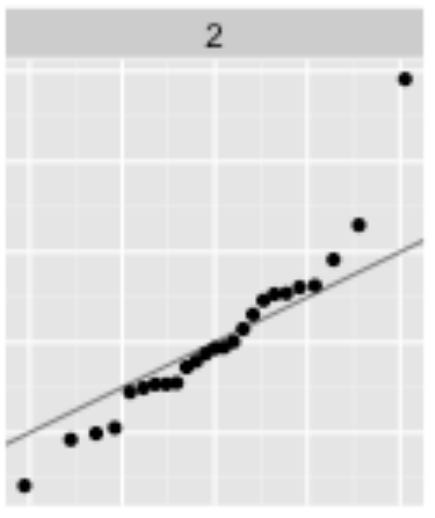}} \end{center} \end{minipage} &Which of these plots is most different from the others? & DF\\ 

\hline 
\end{tabular} 
\label{tbl:visual_stat} 
\end{table*} 

There have been ten experiments conducted using Amazon's Mechanical Turk~\citep{turk} (MTurk) that are being used to evaluate the effects of observers' demographic factors on the inference. Table \ref{tbl:visual_stat} summarizes these experiments. Each of these experiments collected demographic details of the subjects. Experiments 5, 6 and 7 are used to study the learning trend of the observer. The design of experiment 9 incorporated components that allows position of the actual data plot in the lineup, and the sample of nulls, to be evaluated. Section \ref{sec:factor_performance} discusses human factors that may affect the performance of the observer. Section \ref{sec:exp_design} describes the methods used to assess the effects, and Section \ref{sec:result_socio} describes the results.


\section{Factors Potentially Affecting Visual Inference} \label{sec:factor_performance} 

Because visual statistical inference relies on plotting choices, lineup design and human evaluation, it is important to understand how these factors that might affect results.  This section provides a brief description of the factors that are expected to have some impact on the performance of visual statistical inference.  



\begin{itemize} \itemsep 0in
\item{\bf Choice of Visual Test Statistic:}  
There are a variety of ways that data can be plotted and for plots to be enhanced by axes, legends or representations of models. The choice of plot chosen as the visual test statistic, primarily should be appropriate for the problem being investigated. For example, for studying the association between two quantitative variables a scatterplot is usually optimal. Enhancements, such as a line overlaying a scatterplot to represent a fitted model, might be appropriate for studying models for the data.  Some choices of visual test statistics provide better results than others when studying the same problem, which is discussed in \cite{heike:2012}.  This makes sense, because some plot types will make it easier for certain structures to be seen. 

Table \ref{tbl:visual_stat} shows the large variety of visual test statistics used in the Turk studies completed so far. These experiments were conducted to examine many different visual tasks. Side-by-side boxplots were used in experiment 1 to study the importance of a categorical variable in a regression model. For both experiments 2 and 3, a scatterplot is the visual test statistic, but a regression line fitted through the points is overlaid for experiment 2. The experiments studied slightly different problems, for which the difference in plot type was important. Experiment 4 compared polar coordinates to euclidean coordinates on the reading of structure in a barchart.  Experiment 5 examined the style of plot for a bivariate data problem, by comparing results when using a scatterplot, side-by-side dot plots, overlaid densities or side-by-side boxplots. Experiment 6 studied the difference between variations of boxplots. Experiment 7 studied the effects of high dimension low sample size on the separation of groups in low-dimensional projections. The sine illusion, where people read the strength of differences between curves as different when there is a strong sinusoidal structure, is examined in experiment 8. Experiment 9 was conducted to test if there was any structure in an RNA-seq data set measured on soybeans, and used interaction plots classically constructed to examine interaction effects in $2\times 2$ factorial designs. Experiment 10 used different sorts of residual plots for hierarchical linear models as the visual test statistics. 

In experiments 4, 5 and 6, different visual test statistics were used on the same data, providing the ability to examine the effect of these choices. Results are published in \cite{heike:2012} and \cite{XXX}(The violin plot study needs to be referenced). 

\item{\bf Signal in the Data:}
A visual test statistic is chosen for a particular task. If it is well-designed then departures from the null hypothesis will be visible in the plot, especially if the signal is strong. An important factor that helps an observer to identify the actual data plot in a lineup is the strength of the signal in the data. Simulated data was generated for all of the experiments, except 9, enabling the study of signal strength and actual data plot detection. Primarily the results were as expected, that as the signal strength increased observers more frequently picked the actual data plot as different from the null plots. 



\item{\bf Question Design:} A primary purpose of visual inference is to preclude the necessity to pre-specify discoverable features in data that is required by classical statistical inference. Visual inference supports discovery of structure, enabling the the unexpected to surprise us.  This is tightly connected to the task that the human observer is asked to perform. For most purposes the observer is asked a very general question, such as to pick the plot that is most different from the others, and explain the reasons why they see it as different. By being general, all possible discoverable structures are included in the significance calculations. (\citet{buja:2009} has a very complete explanation.)

In the initial Turk experiments (1,2,3) conducted with simulated data to compare the results from visual inference with those obtained by classical inference, the questions posed to the observers were more targeted. This was important in order to make a comparison of the results. For latter studies (4, 7, 10), question posed to the observer was broader. Table \ref{tbl:visual_stat} lists the questions used for each experiment. 

\item{\bf Demographics:}  The experiments collected data on age, gender, education level and geographic location. Each observer self-reported their gender, age in roughly five year intervals, education level as high school or less, some college courses, an undergraduate degree, some graduate courses or a graduate degree. The IP address of the computer afforded the geographical location of the subject. The purpose of collecting this information with each experiment is to examine the effect that they have on the results of visual inference -- ideally very little. 

\item{\bf Learning Trend:} One might expect that as an observer evaluates more lineups they may become more skillful in their evaluations. Each subject in the experiments 5, 6 and 7 evaluated a block of of ten lineups of the same type of data plot. The ten lineups were randomly chosen from the lineups  produced for each study. Before evaluating the first lineup, the observer needs to read instructions and become accustomed to the type of plot used in the lineup. In subsequent lineups being evaluated the type of plot is the same, level of difficulty will be different. It is possible that the observer becomes more skillful at recognizing the most different plot, either by more often detecting the actual data plot or more quickly reporting their choice. These two ways of measuring learning trend are evaluated on data collected from experiments 5, 6 and 7.




\item{\bf Location of Actual Data Plot in the Lineup} For all of the lineups used in the experiments a $5\times 4$ grid of 20 points is used. A random number generator is used to determine the position where the actual data plot is placed in each lineup. With the help of eye-tracking equipments \citet{zhao:2012} tracked the observers' eyes to see how they searched the plots in a lineup. The results suggest that many people read lineups from left to right direction, some read from upward to downward. There is some concern that if the actual data plot is located in the top left of the lineup, observers may find it more easily than if it were placed in the bottom right. Ideally this does not happen. Experiment 9, conducted to test for structure in a real data set, enables this to be studied. Five different locations in the lineup were used to assess how fast and accurately observers identified the actual data plot.




\item{\bf Sample of Null Plots:} In classical inference, the test statistic is compared with the full null distribution, to determine if it is extreme or not. In visual inference the actual plot is compared with plots of a finite number of samples from the null distribution. In the lineups used in the Turk experiments, there are 19 null plots upon which the comparison with the actual data plot is made. These null samples are random draws, and there is a chance that one or more null plots in the lineup may be similar or even more extreme than the actual data plot. The sample of null plots can affect the observer's decision. This is tested with the data from experiment 9, where the experiment was set up with lineups made from different null plots.


\item{\bf Individual Skill or Ability:} Each person may have different visual skill sets, and aptitude for reading statistical plots. This can be examined by having the same lineups evaluated by multiple subjects, so that it can be seen if some subjects consistently detect the actual data plot more often than others.  
\end{itemize}

\section{Experimental Methods}\label{sec:exp_design}

Two of the factors, signal in the data and individual abilities, were studied in \cite{majumder:2013}. The choice of visual test statistic was examined in \cite{heike:2012}. In each of these analyses demographic factors were given a cursory glance, to ensure that they did not have large effects on the results. The design of experiments 5, 6 and 7 enables the examination of learning trend, which is studied in this paper. Experiment 9 was a real test case for visual statistical inference, and in order to understand the significance of the structure in the genomic data, multiple lineups were made in which location of the actual data plot, and the sample of nulls, were randomized. This enables the assessment of the effect of these factors on the results. This section describes the experimental methods used to examine the effects of demography, placement of the actual data plot, sample of null plots and the existence of a learning trend. 



\subsection{Demographic Factors}  

All the ten experiments shown in Table \ref{tbl:visual_stat}, had demographic information collected from subjects, as follows:

\begin{enumerate} \leftmargin 5cm  \itemsep 0in
\item Age group, with categories set to be 18-25, 26-30, 31-35, 36-40, 41-45, 46-50, 51-55, 56-60, above 60.
\item Gender, male or female.
\item Education level, with levels being high school or less, some under grad course, under graduate degree, some graduate courses, and graduate degree.
\item Geographical location, collected from the IP address of the participants' computer, as latitude, longitude, city and country. 
\end{enumerate}


Let $Y_{ij}$ denote the response from observer $i$ on a lineup $j$, with $Y_{ij} =1$ if the actual data plot is chosen, otherwise $Y_{ij} =0$. The factors are examined in association with the observer's response using a mixed effects linear model: 
\begin{equation} \label{eqn:demographic_response}
g( \pi_{ij} )= \mu + \alpha_{k(i)} + \gamma_{l(i)} + \tau_{m(i)}+ \kappa_{s(i)} + \ell_j,  
\end{equation}
where $\pi_{ij}=  E(Y_{ij})$ is the probability that  observer $i$ picks the actual data plot from lineup $j$, 
 $\mu$ is an overall average probability for picking out the data plot from a lineup, $\alpha$, $\gamma$, $\tau$ and  $\kappa$ are the effects of age group $k(i)$, gender $l(i)$, education level $m(i)$ and country name $s(i)$ respectively for observer $i$. The term $\ell_j$ is a random intercept predicting lineup difficulty level and we assume  $\ell_j \sim N(0, \sigma_\ell^2)$, and $g(.)$ denotes the {\it logit} link function $g(\pi)=\log(\pi) - \log(1-\pi); 0 \le \pi \le 1$.

In addition, a model is constructed with time taken as the response variable. Let $Z_{ij}$ denotes the logarithm of time taken for an observer $i$ to evaluate a  lineup $j$. Let $\mu_{ij}=  E(Z_{ij})$ be the average of log(time taken) by  observer $i$ to pick the data panel from lineup $j$. We model this in a mixed effects model of the form
\begin{equation} \label{eqn:demographic_time}
Z_{ij} = \mu + \alpha_{k(i)} + \gamma_{l(i)} + \tau_{m(i)}+ \kappa_{s(i)} + \ell_j+ \epsilon_{ij},  
\end{equation}
where $\mu$ represents overall average of log time taken by an observer to evaluate a lineup, $\alpha$, $\gamma$, $\tau$ and  $\kappa$ are as described in model \eqref{eqn:demographic_response}, $\ell_j$ is a lineup-specific random effect for the time needed to evaluate a lineup, with $\ell_j \sim N(0, \sigma_\ell^2)$ and the overall error $\epsilon_{ijk} \sim N(0, \sigma^2)$.  

\subsection{Learning Trend} Learning trend of a subject can be observed in terms of performance over successive feedbacks received when multiple lineups are shown for evaluation. Experiments 5, 6 and 7 were used for this. Each subject was shown a total of 10 lineups randomly selected from a pool of lineups. The lineups are not necessarily with the same difficulty levels, but the order of lineups was randomized. The responses of the lineups were recorded by attempt 1 through 10. Attempt 1 means that the response is for the first lineup the observer evaluates and attempt 10 refers to the response for the 10th lineup. The goal is to estimate whether performance of the observer improves, or changes, from attempt 1 to attempt 10. 

It should be noted that we are examining the observer's performance, when we model response as detected or not, but this is not the goal of visual inference. Visual inference is constructed to measure the significance of structure discovered in data. It is expected that some observers will be more skilled at reading data plots, and hence, more readily detect the plot that is different. It is also expected that as observer's gain experience in evaluating lineups that they become more proficient in reading data plots, particularly if feedback is given on whether the actual data plot was chosen or not. Choosing the actual data plot will be more difficult in some lineups than others, and indeed should happen purely by chance in some lineups.
So in this context, detected, or not, is used as a response to examine individual differences. 


Let $Y_{ijk}$ denote the response from observer $i$ on lineup $j$ at their $k^{th}$ evaluation attempt, where $Y_{ijk}=1$ if the observer detected the actual data plot otherwise  $Y_{ijk}=0$. Let $\pi_{ijk}=  E(Y_{ijk})$ be the probability that  observer $i$ picks the actual data plot from lineup $j$ in their $k^{th}$ attempt. Learning trend is assessed using a generalized mixed effects model of the form
\begin{equation} \label{eqn:trend_response}
g( \pi_{ijk} )= \mu + \alpha_k + u_i +  a_{i} k + \ell_j,  
\end{equation}
where $\mu$ is an overall average probability for picking out the data plot from a lineup, $\alpha_k$ is the effect of the $k^{th}$ attempt on the probability, with $\alpha_1 = 0$ and $k = 1, ..., K$, $u_i$ and $a_i$ are observer specific random effects, $i = 1, ..., I$. The term, $u_i$ is a random intercept, describing a basic subject-specific ability, with $u_i \sim N(0, \sigma_u^2)$. 
The term $a_i$ is a random slope capturing an individual's specific learning effect over the course of $K$ attempts, where $a_i \sim N(0, \sigma_a^2)$. 
For $\ell_j$ a normal distribution, $N(0, \sigma_\ell^2)$, is assumed, and $\ell_j$ is a random intercept predicting lineup difficulty level. $g(.)$ denotes the {\it logit} link function $g(\pi)=\log(\pi) - \log(1-\pi); 0 \le \pi \le 1$. The inverse link function, $g^{-1}(.)$, from equation \ref{eqn:trend_response} leads to the estimate of the subject and the lineup specific probability of successful evaluation in $k^{th}$ attempt by a single observer as 
\begin{equation} \label{eqn:trend_power}
\hat p_{ijk} =  g^{-1}(\hat{\mu} + \hat{\alpha}_k + \hat{u}_i +  \hat{a}_i k + \hat{\ell}_j).
\end{equation}

When time taken to evaluate a lineup is used as the response, let $Z_{ijk}$ denote the logarithm of time taken for an observer $i$ to evaluate a  lineup $j$ in his/her $k^{th}$ attempt. Let $\mu_{ijk}=  E(Z_{ijk})$ be the average of log(time taken) by  observer $i$ to pick the data panel from lineup $j$ in his/her $k^{th}$ attempt. We model this in a mixed effects model of the form
\begin{equation} \label{eqn:trend_time}
Z_{ijk} = \mu + \alpha_1 + \alpha k + u_i +  a_{i} k + \ell_j + \epsilon_{ijk},  
\end{equation}
where $\mu$ represents overall average of log time taken by an observer to evaluate a lineup, $\alpha$ is the average change in log time taken for each additional attempt,  $\alpha_1$ is an offset in log time taken for the first attempt. All other effects are random effects: $u_i$ is a subject-specific intercept representing individual speed of an observer with $u_i \sim N(0, \sigma_u^2)$, $a_i$ is a subject-specific slope representing the deviation of the speed-up (or -down) by attempt $k$, with $a_i \sim N(0, \sigma_a^2)$, $\ell_j$ is a lineup-specific random effect for the time needed to evaluate a lineup, $\ell_j \sim N(0, \sigma_\ell^2)$ and the overall error $\epsilon_{ijk} \sim N(0, \sigma^2)$.
Equation \ref{eqn:trend_time} leads to the estimate of the subject and the lineup specific time taken for an evaluation in $kth$ attempt by a single observer as 
\begin{equation} \label{eqn:trend_time_est}
\hat \mu_{ijk} =  \hat{\mu} + \hat{\alpha_1}+ \hat{\alpha}k + \hat{u}_i +  \hat{a}_i k + \hat{\ell}_j.
\end{equation}

To fit all these mixed effect models the function $lmer()$ is used from R package $lme4$ by \cite{lme4:2011}. To obtain the $p$-values of fixed effect parameters estimates, normal approximation is used for $Z$ scores computed as the ratio of estimates to the estimated standard errors.  

\subsection{Location Effect} \label{sec:location_design} Experiment 9 studied significant expression in an RNA-seq study, and was designed so that location effect of the actual data plot in a lineup could also be assessed. The data used, documented in \cite{atwood:2013}, measures gene expression of soybean by RNA-seq methods. Two factors were of primary interest: genotype (call them RPA and EV) and the interaction between genotype and iron condition (sufficient or insufficient). The $2\times2$ factorial design provides four treatment conditions. For each gene there were 11 data points, because each treatment contained two or three replicates.  
In large studies such as this there is a valid question whether the data exhibits any structure at all, or if the small $p$-values are simply occurring by chance, from the massive multiple testing. This overall significance is studied by visual inference. 

For the study, the routine processing of the data was conducted to find (1) the gene that exhibited the most significance difference in expression between genotypes, (2) the gene that exhibited the most different expression on RPA, but insignificant expression on EV. Two different types of plots would be made to examine this. For (1) side-by-side dotplots of the expression levels for the one gene by genotype were created. For (2) an interaction plot, with iron sufficiency represented on the horizontal axis, expression vertically, and genotype by color. Colored lines connect the means for the two genotypes over iron treatment. The null plots were created by randomizing the experimental design, and re-doing everything else as done with the actual data. Any genes found under these circumstances had expression differences that can only be considered to by consistent with random differences. Full results are documented in \cite{tengfei:2013}. 

Multiple lineups of the data were produced. The actual plot was randomly placed in five different locations in a lineup of size 20. The locations were  1, 8, 12, 17, 20 for genotype effect (1) and 2, 9, 12, 16, 20 for interaction effect (2). Five different sets of null plots were used to produce 5 lineups for each location position, creating a total of 25 lineups for studying the genotype effect, and another 25 lineups for studying the interaction effect. Each observer saw three lineups, one for genotype, one for interaction, and one easy lineup that was used to help clean the data. 

To examine if the difference in detection rate among the locations is statistically significant a one-way multivariate analysis of variance (MANOVA) model is fit to the data.
Let $\mathbf{Y}=(Y_1,Y_2, ... , Y_p)^\top$ be a vector of random variables with dimension $p (=5)$, the total number of null sets, and let $\mathbf{Y}_{ij}$ represents $jth$ vector response for $ith$ location with $i=1,2, ..., I(=5)$. Because the same data plot is shown in each lineup, it is assumed that there could be some association between the responses for each null set, which suggests the MANOVA model rather than a univariate ANOVA.  The MANOVA model 
\begin{equation}\label{manova}
\mathbf{Y}_{ij} = \mu_{i} + \epsilon_{ij}
\end{equation}
where $\mu_{i}= (\mu_{1i},\mu_{2i}, ..., \mu_{pi})^\top$ is the mean vector for location $i$ and $Var(\epsilon_{ij})=\Sigma$, tests for significant difference between the means. 

\subsection{Data Collection Methods}  Human subjects were recruited to evaluate the experimental lineups through MTurk \citep{turk}.  It is an online work place where people from around the world can perform tasks, that computers would have difficulty doing, and get paid. Usually tasks are very simple and no specialized training is required to do them. Tasks are designed for anyone to do but some tasks may require some skills depending on the recruiters' need. Each task is usually planned to completed in a short time.  The amount of money paid for each task is very small, on the order of minimum wage in the USA. 



We designed and developed a web application which enables the display of lineups to the observers as per experimental need. The MTurk workers were redirected to this web application to complete their assigned tasks. The responses were collected, stored automatically into a local database server, along with demographic information, age group, gender and education level. The time taken for each evaluation is computed based on the time the plot was shown and the time the feedback was received. It is measured in seconds. The location of the observer is determined by the ip address of the observer.

\section{Results}\label{sec:result_socio}

\subsection{Overview of the Data} 
\begin{figure}[htbp] 
   \centering
   \includegraphics[width=4.5in]{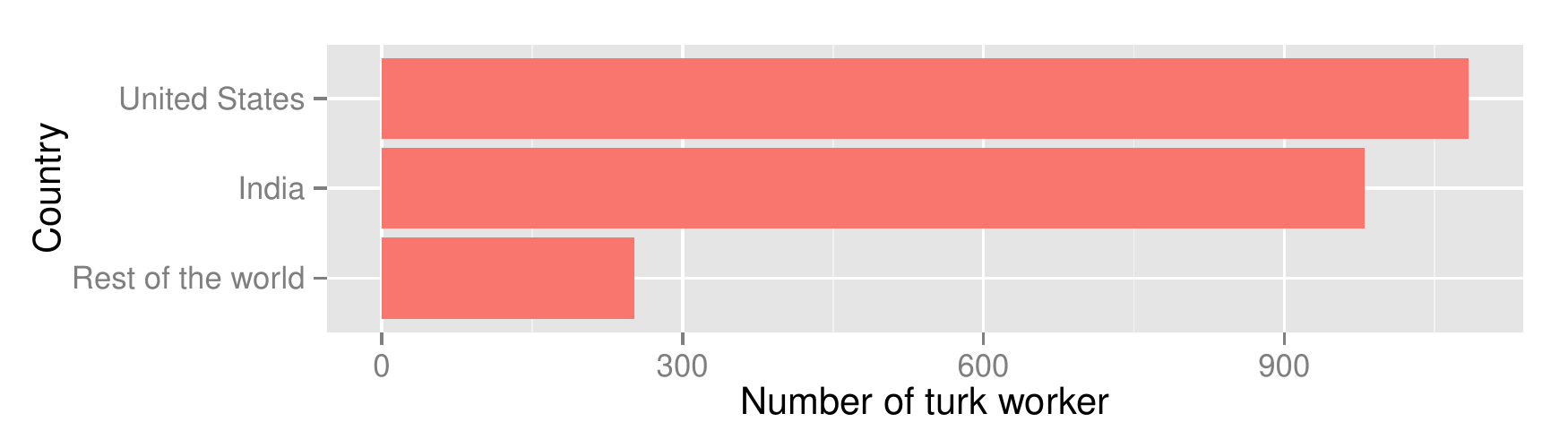} 
   \includegraphics[width=4.5in]{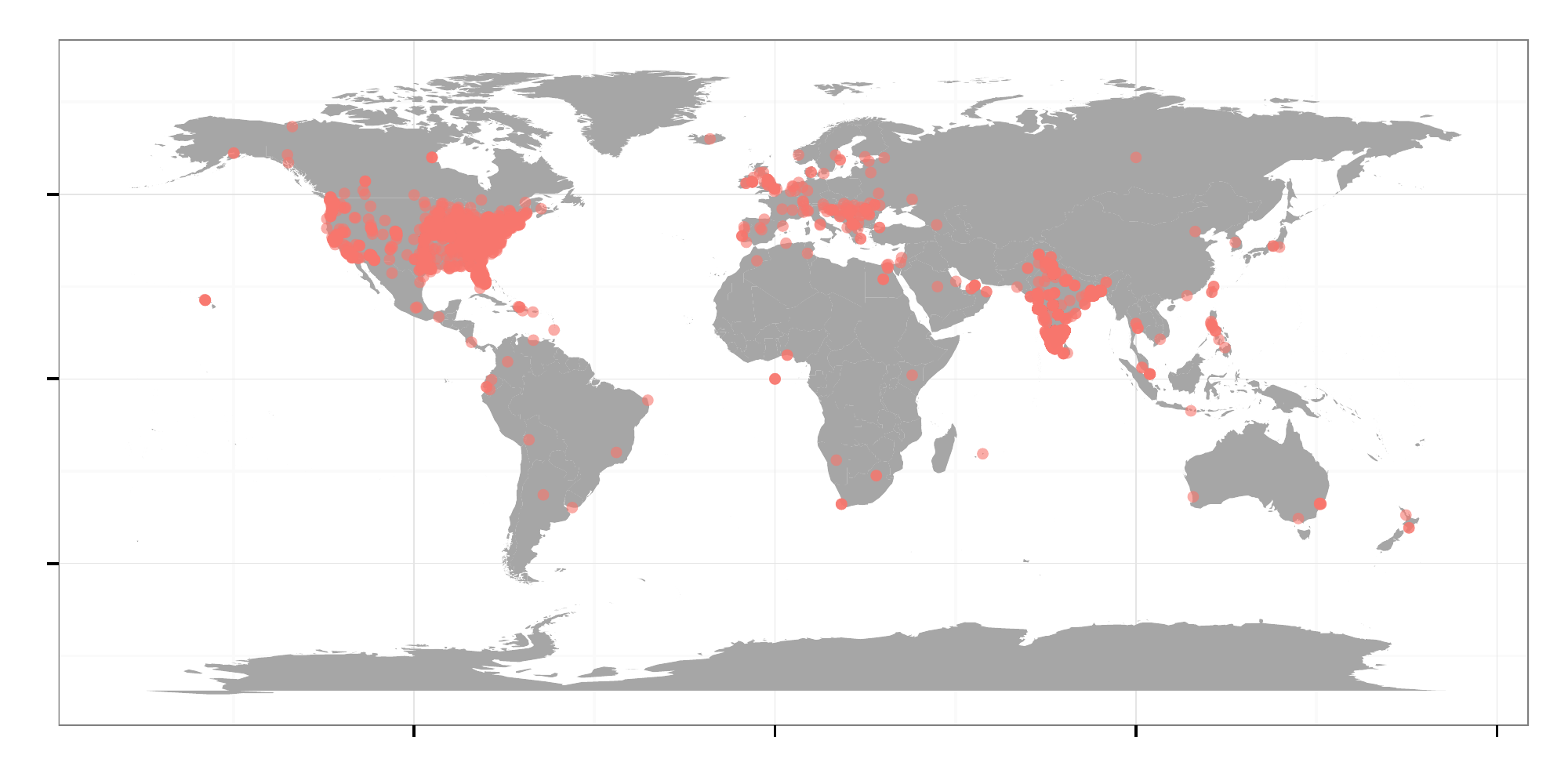}    
   \caption{Location of the Amazon Mechanical Turk workers participating our study. Most of the people are coming from India and United States even though there are people from around the world.}
   \label{fig:turker_location}
\end{figure}

A total of 2321 participants provided feedback data on the lineups in ten different experimental studies. Figure \ref{fig:turker_location} displays the locations of participants around the world.  Most of the participants were from the United States and India. There were 76 other different countries represented. This provides a diverse pool of participants. The diversity in not only in geographic but also in their gender, age group and education level as can be seen in Table \ref{tbl:demographics}.  It is interesting that there were large number of female participants even though there were lot of participants worked from developing countries.

Besides United States and India, countries such as Canada, Romania, United Kingdom and Macedonia have more than 10 participants each. The rest of the 70 countries have less than 10 participants. The distribution of participants remains similar in all ten experiments. That may be due to the time of day when the experiment was made available. 

\begin{table}[hbtp]
\caption{Demographic information of the subjects participated the MTurk experiments. Average time taken for evaluating a lineup is shown in seconds.}
\centering
\scalebox{.85}{
\begin{tabular}{rlrrcr}
  \hline
  && \multicolumn{2}{c}{Participants}&Average& Number of\\
  \cline{3-4}
 Factor & Levels & Total &\% & Time & Responses \\ 
  \hline
Gender & Male & 1348&57.63 & 48.51 & 13493 \\ 
   & Female & 991&42.37 & 43.75 & 10564 \\ 
\hline
  Education & High school or less & 193 &8.24& 37.21 & 2241 \\ 
   & Some under graduate courses & 418 & 17.85&42.84 & 4070 \\ 
   & Under graduate degree & 584 &24.93& 44.29 & 5775 \\ 
   & Some graduate courses & 245 &10.46& 43.43 & 2460 \\ 
   & Graduate degree & 902 &38.51& 52.18 & 9511 \\ 
\hline
  Age & 18-25 & 740 &31.61& 42.97 & 7311 \\ 
   & 26-30 & 547 &23.36& 46.27 & 5585 \\ 
   & 31-35 & 376 &16.06& 44.27 & 3923 \\ 
   & 36-40 & 257 &10.98& 55.03 & 2714 \\ 
   & 41-45 & 141 &6.02& 43.90 & 1519 \\ 
   & 46-50 &  95 &4.05& 49.29 & 1003 \\ 
   & 51-55 &  83 &3.54& 48.67 & 867 \\ 
   & 56-60 &  64 & 2.73& 59.73 & 678 \\ 
   & above 60 & 38 &1.62& 48.67 & 457 \\ 
\hline   
  Country  & United States & 1087 &46.83& 39.64 & 10769 \\  
  & India & 980 &42.22& 52.63 & 10227 \\ 
  & Rest of the world & 254&10.94 & 46.86 & 2819 \\ 
  
   \hline
\end{tabular}
}
\label{tbl:demographics}
\end{table}

%

The largest number of participants are from age group 18 to 25, with the majority between 18 to 35. Interestingly, there are many participants in the older age groups as well. The United States shows a more uniform participation beyond age 30 but Indian participants are primarily younger. 

In terms of academic background, the largest participant group is graduate degree, with a total of 902 participants, about 38.51\%. Examining the location in conjunction with the academic background reveals that there are more Indian participants that list themselves as having a graduate degree. In retrospect this was realized to be a language confusion, as any university degree is considered to be a graduate degree in India.  Unlike the definition in North America where graduate degree means Masters level education. Most of the USA participants list themselves as having an undergraduate degree or at least have some undergraduate courses. 

The distribution of male and female participants are similar among all age groups except for age group 18-25 in India where proportion of female participants is lower. The distribution of education levels also differ slightly across the countries for this age group. 

A total of 1911 lineups were evaluated in the ten experiments. Each person evaluated at least 10 lineups except for experiment 9 where three lineups were evaluated by each person. A test plot was shown to each observer and the feedback received from that plot is used to ensure data quality. Data from participants who got it wrong were removed, and then results for just this lineup were removed for all participants who responded correctly. In some cases the demographic information were not provided by the participants. Also, for some ip address, the actual geographical locations could not be retrieved. This resulted some missing demographic information.

\subsection{Demographic Factors} 


To explore the significance of the demographic factors model \eqref{eqn:demographic_response} is fit to the data, with age, country, education and gender as fixed effects. To estimate the significance of the factors, reduced models are fit with that factor removed from the model. Table ~\ref{tbl:anova_factor} summarizes the results. All of the demographic factors are significantly different in describing the the detection rate, except for gender.  With respect to time taken, all of the demographic factors are significant. 

\begin{table}[htbp]
\centering
\caption{Analysis of variance (ANOVA) table comparing full model, all the demographic factors, with the reduced models, obtained by removing respective factor variable. Gender does not have a significant effect on detection rate, but does on time to respond. All factors significantly affect time to respond.}
\scalebox{.9}{
\begin{tabular}{rlrrrrrr}
  \hline
Model & & AIC & BIC & logLik & Chisq & Chi.Df & $p$-value \\ 
  \hline
Proportion Correct &Full & 23822.00 &23943.00 &-11896.00  & &  &  \\ 
& \multicolumn{2}{l}{Reduced}  &  &  &  & \\
  &Age & 23835.22 & 23907.93 & -11908.61 & 25.50 & 6 & $<$0.001 \\ 
  &Country  & 24099.57 & 24204.72 & -12036.78 & 281.85 & 2 & $<$0.001 \\ 
  &Education  & 23837.48 & 23926.34 & -11907.74 & 23.77 & 4 & $<$0.001 \\ 
  &Gender  & 23821.88 & 23934.98 & -11896.94 & 2.17 & 1 & 0.140 \\ 
   \hline
   Log Time &Full &  51904.00 &52034.00 & -25936.00 & &  &  \\ 
& \multicolumn{2}{l}{Reduced}  &  &  &  & \\
&Age  & 52435.63 & 52516.41 & -26207.81 & 543.18 & 6 & $<$0.001 \\ 
 & Country  & 52849.92 & 52963.15 & -26410.96 & 949.47 & 2 & $<$0.001 \\ 
  &Education  & 52012.68 & 52109.61 & -25994.34 & 116.23 & 4 & $<$0.001 \\ 
  &Gender  & 51970.57 & 52091.74 & -25970.28 & 68.12 & 1 & $<$0.001 \\ 
\hline
\end{tabular}
}
\label{tbl:anova_factor}
\end{table}

Parameter estimates from the model fits are shown in Table \ref{tbl:model_result_demographics}. The first level of each factor serves as the baseline for the model. The age group 36-40 significantly has a higher detection rate than the 18-25, and to a lesser extent this is also true for age groups 31-35 and above 50. Participants from India have a similar detection rate to those from the USA, but ones from the rest of the world had a significantly improved detection rate. Subjects who listed themselves as having a graduate degree had a significantly higher detection rate. For time to respond, all age groups were significantly slower than the 18-25 year olds.  Participants from India were slower, and those from the rest of the world were faster than Americans. A higher degree corresponded to longer time, and males were slower than females!

\begin{table}[htbp]
\centering
\caption{Parameter estimates of models \eqref{eqn:demographic_time} and \eqref{eqn:demographic_response} fitted for average log time taken and detection rate, respectively. For time taken all the demographic factors are significant. For detection rate, age group 36-40, rest of the world and graduate degree are significantly different. For gender no difference in performance is observed. Lineup variability is estimated to be very large for model \eqref{eqn:demographic_response}.}
\scalebox{0.86}{
\begin{tabular}{rlrrrrrrrrrr}
  \hline
Demographic&& \multicolumn{4}{c} {Log Time (model \ref{eqn:demographic_time})} & &\multicolumn{4}{c} {Detection rate (model \ref{eqn:demographic_response})} \\

\cline{3-6} \cline{8-11} 
Factor & Level& Est & SE & Zval & $p$-value && Est & SE & Zval & $p$-value  \\ 
  \hline
\multicolumn{2}{l}{Fixed Effect} &  &  &  &  &  & & && \\ 
&$\mu$ & 3.360 & 0.013 & 249.21 & $<$0.001 &   & -0.683 & 0.071 & -9.64 & $<$0.001 \\ 
\hline
Age ($\alpha$)& 18-25&0.000   &---  &---  &---  &  &--- &--- &---& ---\\   
 &26-30 & 0.058 & 0.013 & 4.50 & $<$0.001 &   & 0.062 & 0.049 & 1.27 & 0.206 \\ 
  &31-35 & 0.068 & 0.014 & 4.72 & $<$0.001 &   & 0.115 & 0.055 & 2.08 & 0.038 \\ 
  &36-40 & 0.231 & 0.016 & 14.05 & $<$0.001 &   & 0.310 & 0.063 & 4.93 & $<$0.001 \\ 
  &41-45 & 0.176 & 0.021 & 8.56 & $<$0.001 &   & 0.158 & 0.081 & 1.96 & 0.050 \\ 
  &46-50 & 0.272 & 0.024 & 11.29 & $<$0.001 &   & 0.141 & 0.096 & 1.47 & 0.143 \\ 
  &above 50 & 0.352 & 0.018 & 19.19 & $<$0.001 &   & 0.147 & 0.071 & 2.06 & 0.039 \\ 
  \hline
Country($\kappa$)& United States&0.000   &---  &---  &---  &  &--- &--- &---& ---\\  
  &India & 0.101 & 0.011 & 9.11 & $<$0.001 &   & 0.058 & 0.043 & 1.33 & 0.183 \\ 
  &Rest of world & -0.129 & 0.009 & -13.82 & $<$0.001 &   & 0.185 & 0.035 & 5.22 & $<$0.001 \\ 
  \hline
Education($\tau$) &High school or less&0.000   &---  &---  &---  &  &--- &--- &---& ---\\     
 &Under grad courses & 0.042 & 0.013 & 3.25 & 0.0011 &   & -0.083 & 0.050 & -1.65 & 0.098 \\ 
  &Under grad degree & -0.037 & 0.012 & -3.21 & 0.0013 &   & -0.044 & 0.045 & -0.97 & 0.331 \\ 
  &Graduate courses & 0.117 & 0.013 & 9.12 & $<$0.001 &   & 0.070 & 0.050 & 1.42 & 0.157 \\ 
  &Graduate degree & 0.046 & 0.011 & 4.12 & $<$0.001 &   & 0.182 & 0.043 & 4.22 & $<$0.001 \\ 
  \hline
Gender ($\gamma$)& Female&0.000   &---  &---  &---  &  &--- &--- &---& ---\\      
  &Male & 0.078 & 0.009 & 8.26 & $<$0.001 &   & 0.055 & 0.036 & 1.50 & 0.133 \\ 
  \hline
\multicolumn{2}{l}{Random Effect} &  &  &  &  &  & & && \\ 
& lineup($\sigma_{\ell}$) & 0.082 & 0.287 &  &  &   & 5.259 & 2.293 &  &  \\ 
 & Error($\sigma$) & 0.479 & &  &  &  & \mm{? 0.692}&& &\\ 
   \hline
\end{tabular}
}
\label{tbl:model_result_demographics}
\end{table}

\begin{figure}[htbp] 
   \centering
   \includegraphics[width=4in]{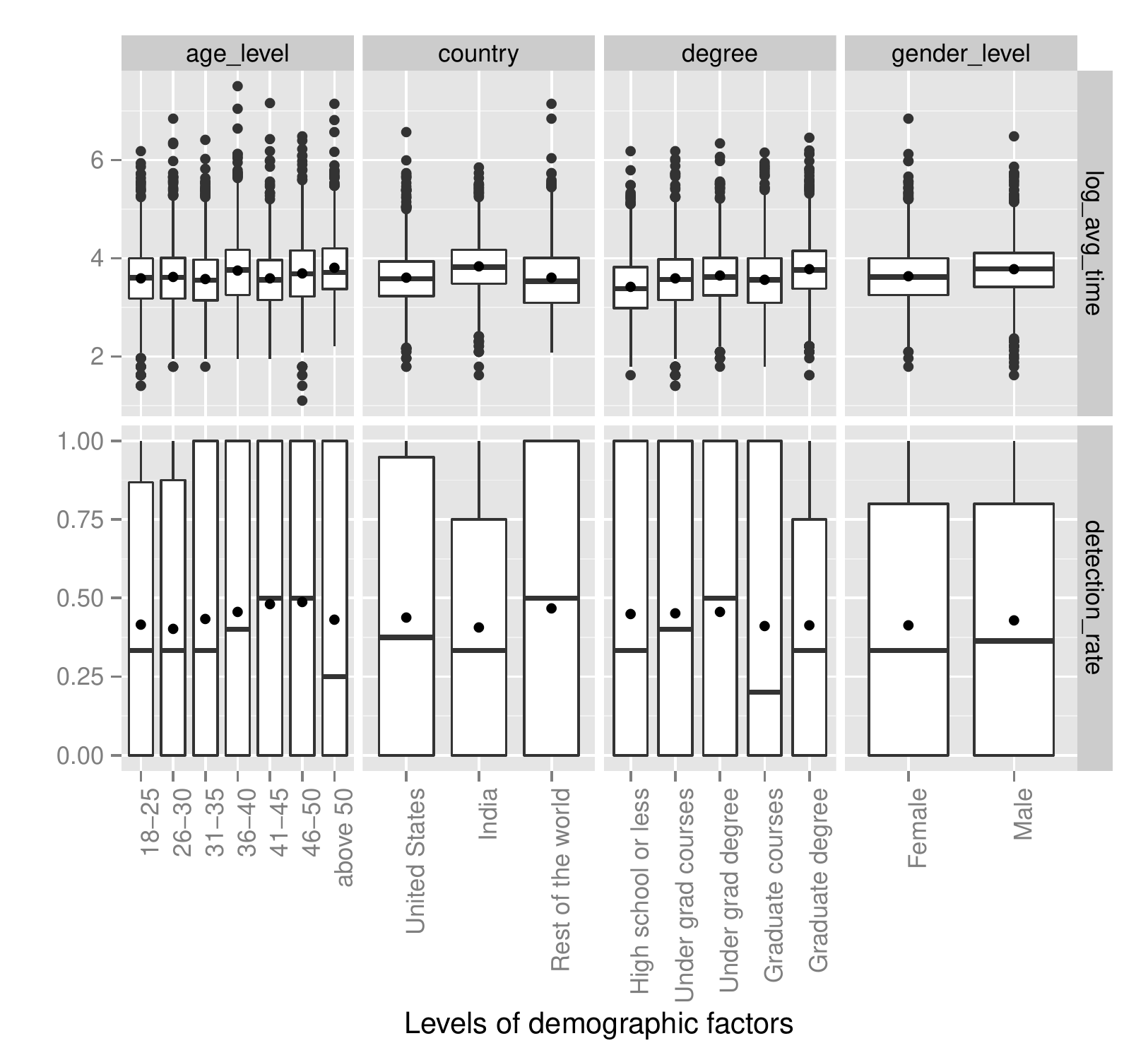} 
   \caption{Boxplots of average log time taken and proportion correct responses (detection rate) of all the lineups plotted for each demographic factor levels. The dots inside the boxes represent means. Some differences in means of various demographic factors are observed. Variability in detection rate indicates large variability in lineup difficulties. }
   \label{fig:demographic_effect}
\end{figure}

Figure \ref{fig:demographic_effect} supports the model results. The average time taken to respond (natural log) and detection rate are computed by lineup for the different demographic factor levels, and displayed using boxplots. Since the models use the mean values, rather than median, these are represented by a dot inside the boxplots. 
The means and medians for response time (top row) are essentially equal. There are some differences between the two in the detection rate plots (bottom row). The differences in these values for different factors, that was tested by the models, can be seen in the plots. However, particularly for detection rate, the variation is huge. A large component of the variation is difficulty of the lineup. In some lineups the actual data plot was distinctively different from the null plots, and we would expect that the detection rate would be close to 100\%. In other lineups there was no difference, making it very difficult to evaluate, with low detection rate and longer time to respond. In Table \ref{tbl:model_result_demographics} notice that lineup specific variance estimate is 5.259 with a standard error of 2.293 which is much higher than the other significant parameter estimates in model \eqref{eqn:demographic_response}. This indicates that the major and most important factor affecting the detection rate is  lineup difficulty.  So although the demographic factors emerge from the model to be statistically significant, the practical significance is minimal. We illustrate this with the following example of graduate degree.



\begin{figure}[htbp] 
   \centering
    \includegraphics[width=3.2in]{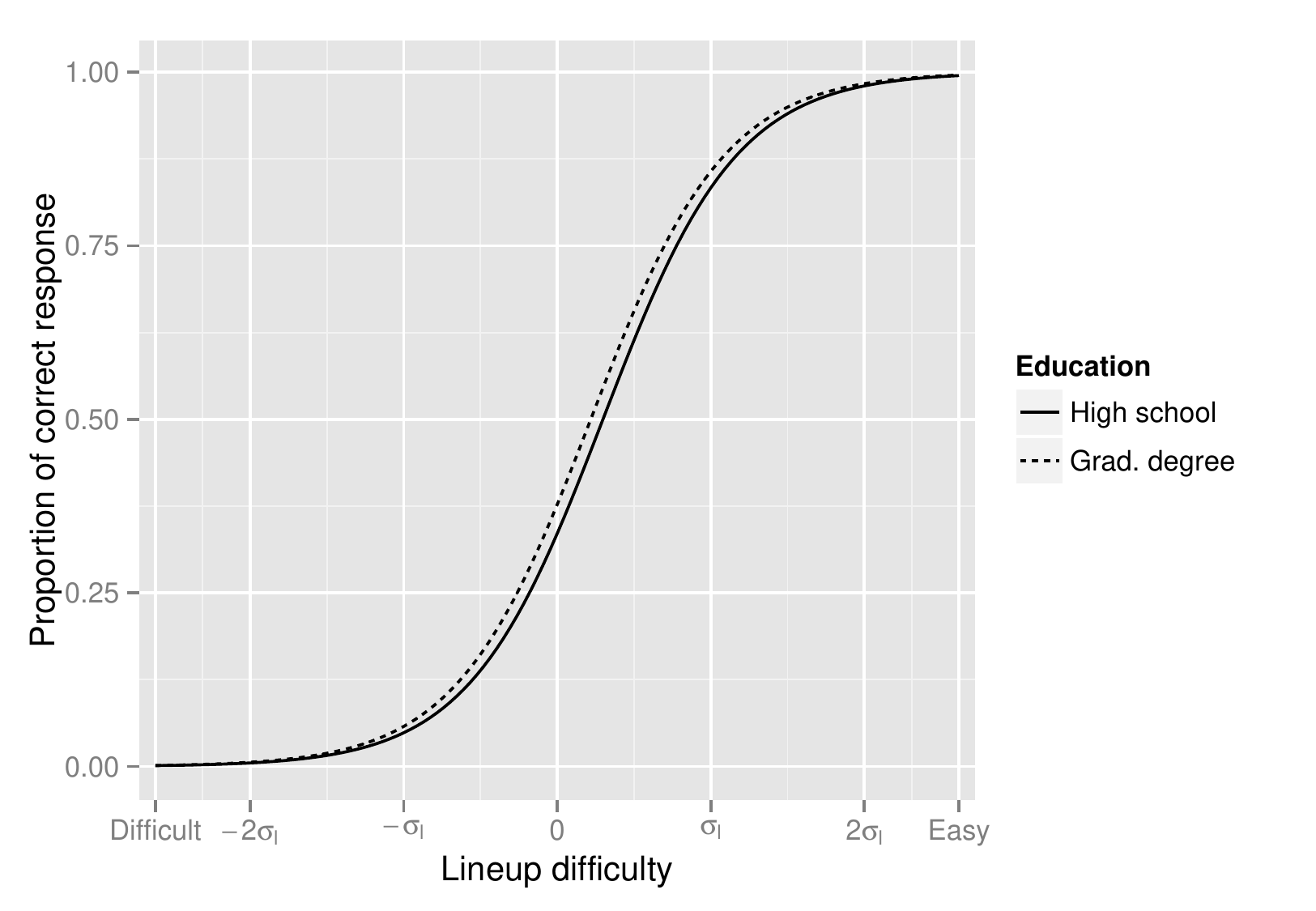} 
    \includegraphics[width=3.2in]{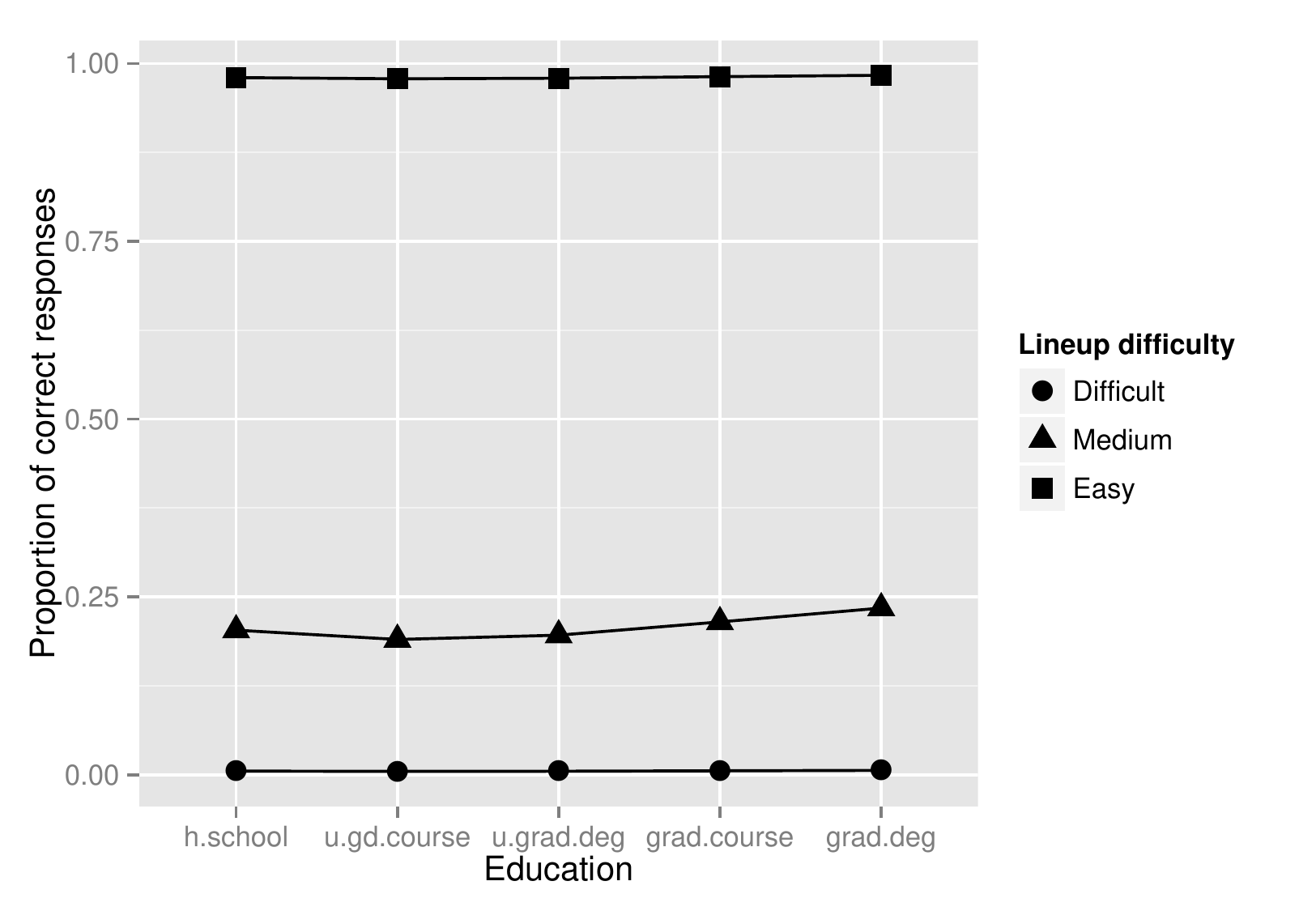} 
   \caption{Proportion of correct responses due to graduate degree as compared to high school degree for an 18-25 year old female in the United States. Even though graduate degree is statistically significant, the largest difference in proportion correct is  0.045 which is very negligible. The difference diminishes as we move away one or two standard deviations ($\sigma_{\ell} =2.293$) 
   of lineup variability.}
   \label{fig:practical_impact_graduate}
\end{figure}

While some of the demographic factors are strongly statistically significant, the main source of variation in detection rate is the lineup difficulty. For example, let's examine the effect of graduate degree. To see just how large the effect is, we examine the change in detection rate for a (hypothetical) 18-25 year old female in the United States, with a graduate degree as compared to a high school degree, for an average difficulty lineup (random effect = 0). Plugging in the relevant quantities to the fitted model gives a difference equal to:
$$ \frac {\exp(-0.683+0.182)}{1+\exp(-0.683+0.182)}-  \frac {\exp(-0.683)}{1+\exp(-0.683)} = 0.377- 0.336 =0.041.$$ 
The person with a high school education on average picks the data plot in 33.5\% of lineups having average difficulty, as compare to 37.6\% if they have a graduate degree. This difference is reduced to 2\% for a lineup with one standard deviation order of magnitude difference in difficulty. For two standard deviations it further reduces to 0.3\%. Thus although there is statistically significant difference in proportion correct for some demographic factors, these are not practically significant differences.  Figure~\ref{fig:practical_impact_graduate} illustrates this example showing fitted models for a US 18-25 female with either a high school education or a graduate degree.  Similar calculations show the same negligible impact of age level 36-40 (0.0533 at most)  and country (0.0424 at most) on the probability of correct response. Thus even though some of the demographic factors are statistically significant, practically, demographics do not substantially influence the results. 

\subsection{Learning Trend} Models \eqref{eqn:trend_response} and \eqref{eqn:trend_time} are fitted to the data from experiments 5, 6, and 7 separately. In this model attempt is fitted as a factor variable to allow for all possible non-linear learning trends.  It should be noted that we also examined an alternative to model \eqref{eqn:trend_response} where attempt was linearly fitted as a continuous covariate, but this also was not significant.

Table \ref{tbl:model_result_response} displays the parameter estimates and $p$-values of fixed effect estimates of model \eqref{eqn:trend_response} examining detection rate. There is some evidence of a learning curve only for experiment 6. There are marginally small $p$-values for most of the attempt levels, and positive parameter estimates, suggesting that more attempts increases detection rate. 


\begin{table}[htbp]
\centering
\caption{Parameter estimates of models \eqref{eqn:trend_response} fitted to data from three different experiments using detection rate as the response to assess learning trend. Attempt number is fitted as a factor to enable modeling any sort of non-linear learning trend. Only experiment 6 shows evidence of a learning trend, with detection rate essentially increasing as attempts increase.}
\scalebox{0.81}{
\begin{tabular}{rrrrccrrrccrrrc}
  \hline
& \multicolumn{4}{c} {Experiment 5} & &\multicolumn{4}{c} {Experiment 6} && \multicolumn{4}{c} {Experiment 7}\\

\cline{2-5} \cline{7-10} \cline{12-15} 
 Effect& Est & SE & Zval & $p$-value && Est & SE & Zval & $p$-value && Est & SE & Zval & $p$-value \\ 
  \hline
Fixed &  &  &  & && &  & & & & & &  & \\ 
$\mu$ & -1.304 & 0.179 & -7.28 & $<$0.001 &   & -0.220 & 0.147 & -1.50 & 0.134 &   & -1.737 & 0.481 & -3.61 & $<$0.001 \\ 
  $\alpha_1$ & 0.000 & --- & --- & --- &   & 0.000 & --- & --- & --- &  & 0.000 & --- & --- & ---  \\ 
  $\alpha_2$ & 0.270 & 0.219 & 1.24 & 0.217 &   & 0.262 & 0.158 & 1.66 & 0.098 &   & -0.456 & 0.385 & -1.18 & 0.237 \\ 
  $\alpha_3$ & -0.178 & 0.226 & -0.79 & 0.432 &   & 0.342 & 0.157 & 2.18 & 0.029 &   & -0.105 & 0.386 & -0.27 & 0.786 \\ 
  $\alpha_4$ & 0.083 & 0.224 & 0.37 & 0.712 &   & 0.358 & 0.159 & 2.26 & 0.024 &   & -0.378 & 0.381 & -0.99 & 0.322 \\ 
  $\alpha_5$ & 0.298 & 0.224 & 1.33 & 0.183 &   & 0.376 & 0.159 & 2.36 & 0.018 &   & -0.107 & 0.385 & -0.28 & 0.781 \\ 
  $\alpha_6$ & 0.042 & 0.231 & 0.18 & 0.857 &   & 0.246 & 0.158 & 1.56 & 0.120 &   & 0.026 & 0.407 & 0.06 & 0.949 \\ 
  $\alpha_7$ & 0.283 & 0.230 & 1.23 & 0.217 &   & 0.160 & 0.159 & 1.01 & 0.314 &   & 0.057 & 0.401 & 0.14 & 0.886 \\ 
  $\alpha_8$ & -0.045 & 0.233 & -0.19 & 0.847 &   & 0.341 & 0.160 & 2.13 & 0.033 &   & -0.003 & 0.394 & -0.01 & 0.994 \\ 
  $\alpha_9$ & -0.195 & 0.232 & -0.84 & 0.400 &   & 0.378 & 0.160 & 2.36 & 0.018 &   & 0.204 & 0.436 & 0.47 & 0.639 \\ 
  $\alpha_{10}$ & 0.513 & 0.228 & 2.25 & 0.024 &   & 0.192 & 0.163 & 1.18 & 0.238 &   & -0.213 & 0.432 & -0.49 & 0.622 \\ 
\multicolumn{2}{l}{Random}   &  &  &  &  & & &&& & & &  & \\ 
  $\sigma^2_a$ & $<$0.001 & 0.017 &  &  &   & 0.001 & 0.034 &  &  &   & 0.027 & 0.163 &  &  \\ 
  $\sigma^2_u$ & 0.720 & 0.848 &  &  &   & 0.815 & 0.903 &  &  &   & $<$0.001 & $<$0.001 &  &  \\ 
  $\sigma^2_{\ell}$ & 2.178 & 1.476 &  &  &   & 2.009 & 1.418 &  &  &   & 10.980 & 3.314 &  &  \\ 
   \hline
\end{tabular}
}
\label{tbl:model_result_response}
\end{table}

\begin{figure}[htbp] 
   \centering
    \includegraphics[width=6.3in]{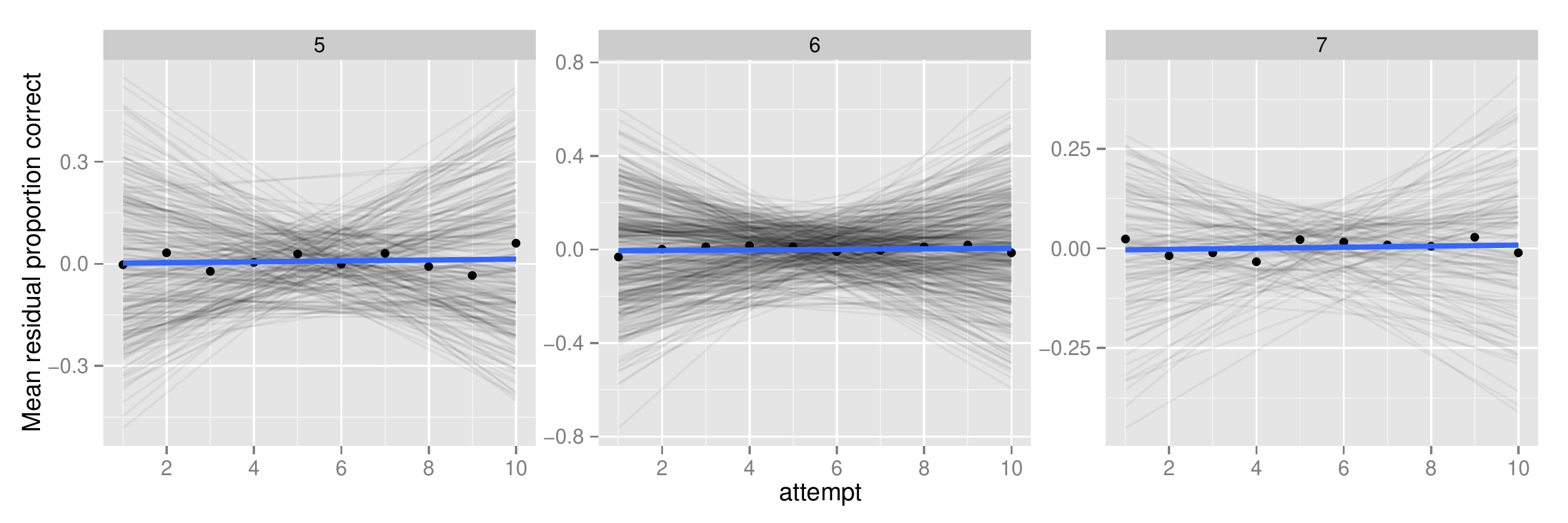} 
   \caption{ Least square lines fitted through the subject specific residual proportion correct obtained from model \eqref{eqn:trend_response} fitted without attempt are plotted against attempt. Subject specific positive and negative slopes are observed. Mean residuals are shown as dots and least square regression lines fitted through the points show no overall learning trend in each of the three experiments.}
   \label{fig:learning_trend_response}
\end{figure}

To visualize how the detection rate effect over successive attempts, we fitted model \eqref{eqn:trend_response} excluding the covariates related to attempt from the model and computed the residuals. Least square regression lines were fitted through the subject specific residuals as shown in Figure~\ref{fig:learning_trend_response}. The averages of these residuals for each of the attempts are shown as dots. Two important features were observed; one is subject specific variability and the other is random slope with attempts which indicates subjects specific learning trend. Some subjects show improvement over time and some show the decrease in performance. Although, the model fit experiment 6 suggested a statistically significant learning curve the plots indicates that is minimal. 

Table \ref{tbl:model_result_time} presents the results of model \eqref{eqn:trend_time} where response time is examined with respect to attempt. Attempt is modelled as a linear covariate here, with a shift factor used to adjust for additional time on the first lineup evaluated. Interestingly, time to respond significantly decreases as number of attempts increases, for all three experiments. 
The parameter $\alpha$ for fixed effect covariate attempt is highly significant in all the experiments. The negative estimates suggest that on an average later attempt took less time for an evaluation. Even though observers did not improve the performance over attempt, they became efficient in responding faster. The parameter $\alpha_1$ for first attempt is also highly significant. The positive estimates of $\alpha_1$ indicates that first attempt made by an observer required much more times than other attempts. It is because for initial attempt the observer might have gone through instructions and became familiar with the experimental environment. Each page in the web site contains information about choice reason and the observers's confidence level. Also, the first page asks for observer Identification to be typed. The later pages of the web site was similar just the lineup was changed. Thus in the later attempts an observer does not need to spend any time for reading instructions. The model reflects that fact.

\begin{table}[htbp]
\centering
\caption{Parameter estimates of model \eqref{eqn:trend_time} fitted for log time taken to evaluate a lineup. Both fixed effect parameters of Attempt ($\alpha_1$ and $\alpha$) are highly significant for all three experiments 5, 6 and 7.}
\scalebox{0.81}{
\begin{tabular}{rrrrccrrrccrrrc}
  \hline
& \multicolumn{4}{c} {Experiment 5} & &\multicolumn{4}{c} {Experiment 6} && \multicolumn{4}{c} {Experiment 7}\\

\cline{2-5} \cline{7-10} \cline{12-15} 
 Effect& Est & SE & Zval & $p$-value && Est & SE & Zval & $p$-value && Est & SE & Zval & $p$-value \\ 
  \hline
Fixed &  &  &  & && &  & & & & & &  & \\ 
$\mu$ & 3.817 & 0.039 & 97.38 & $<$0.001 &   & 3.901 & 0.033 & 118.19 & $<$0.001 &   & 3.731 & 0.054 & 69.04 & $<$0.001 \\ 
  $\alpha_1$ & 0.326 & 0.035 & 9.35 & $<$0.001 &   & 0.335 & 0.029 & 11.40 & $<$0.001 &   & 0.280 & 0.050 & 5.63 & $<$0.001 \\ 
  $\alpha$ & -0.038 & 0.004 & -9.30 & $<$0.001 &   & -0.039 & 0.004 & -10.19 & $<$0.001 &   & -0.029 & 0.007 & -4.22 & $<$0.001 \\ 
\multicolumn{2}{l}{Random}   &  &  &  &  & & &&& & & &  & \\ 
  $\sigma^2_a$ & 0.001 & 0.027 &  &  &   & 0.002 & 0.045 &  &  &   & 0.002 & 0.049 &  &  \\ 
  $\sigma^2_u$ & 0.259 & 0.509 &  &  &   & 0.245 & 0.495 &  &  &   & 0.134 & 0.366 &  &  \\ 
  $\sigma^2_{\ell}$ & 0.008 & 0.091 &  &  &   & 0.040 & 0.199 &  &  &   & 0.055 & 0.235 &  &  \\ 
  $\sigma^2$ & 0.211 & 0.460 &  &  &   & 0.251 & 0.501 &  &  &   & 0.206 & 0.454 &  &  \\ 
   \hline
\end{tabular}
}
\label{tbl:model_result_time}
\end{table}



\begin{figure}[htbp] 
   \centering
 \includegraphics[width=6.3in]{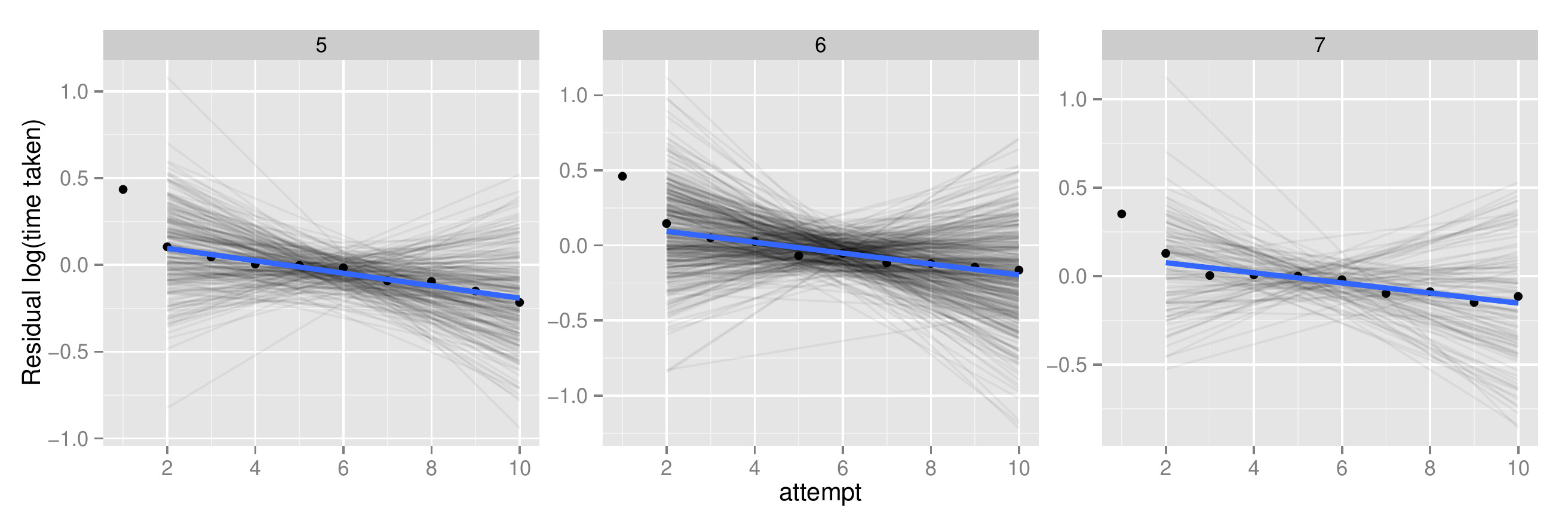}    
   \caption{Least square regression lines fitted through the subject specific residuals obtained by  fitting model \eqref{eqn:trend_time} without covariate attempt. Differences in subject specific slopes are observed. Some of the subjects did worse over successive attempts while others did better. Averages of these residuals are plotted as dots and least square regression lines are fitted to obtain overall trends. For all the three experiments the overall downward slopes are statistically significant which indicates that MTurk workers take less time as they progress through their attempts.}
   \label{fig:learning_trend_time}
\end{figure}

To visualize how the time taken reduces over the successive attempts, we fitted model \eqref{eqn:trend_time} excluding the covariate attempt from the model and computed the residuals. Least square regression lines are fitted through the subject specific residuals. Subject specific slopes are much different in each of the three experiments as we see in Figure \ref{fig:learning_trend_time}. Some subjects improved over attempts by taking less time in the later attempts while others got worse. The averages of these residuals for each attempt are plotted as dots. Least square regression lines are fitted to these points excluding the first attempt since for first attempt we fitted an indicator covariate. The downward trends are evident in the plots. All the slopes are highly significant. As expected we observed large positive residuals for each of the experiments for first attempt.

\subsection{Lineup Design}

A total of 111 subjects were recruited to evaluate lineups designed to investigate the location effect of the actual data plot in the lineup as described in Section~\ref{sec:location_design}. Each subject evaluated two lineups;  one for Interaction effect and the other for Genotype.  In total there were 222 responses on 50 lineups. The data on the test lineup were excluded from the analysis.

\begin{figure}[htbp] 
   \centering
    \includegraphics[width=6.5in]{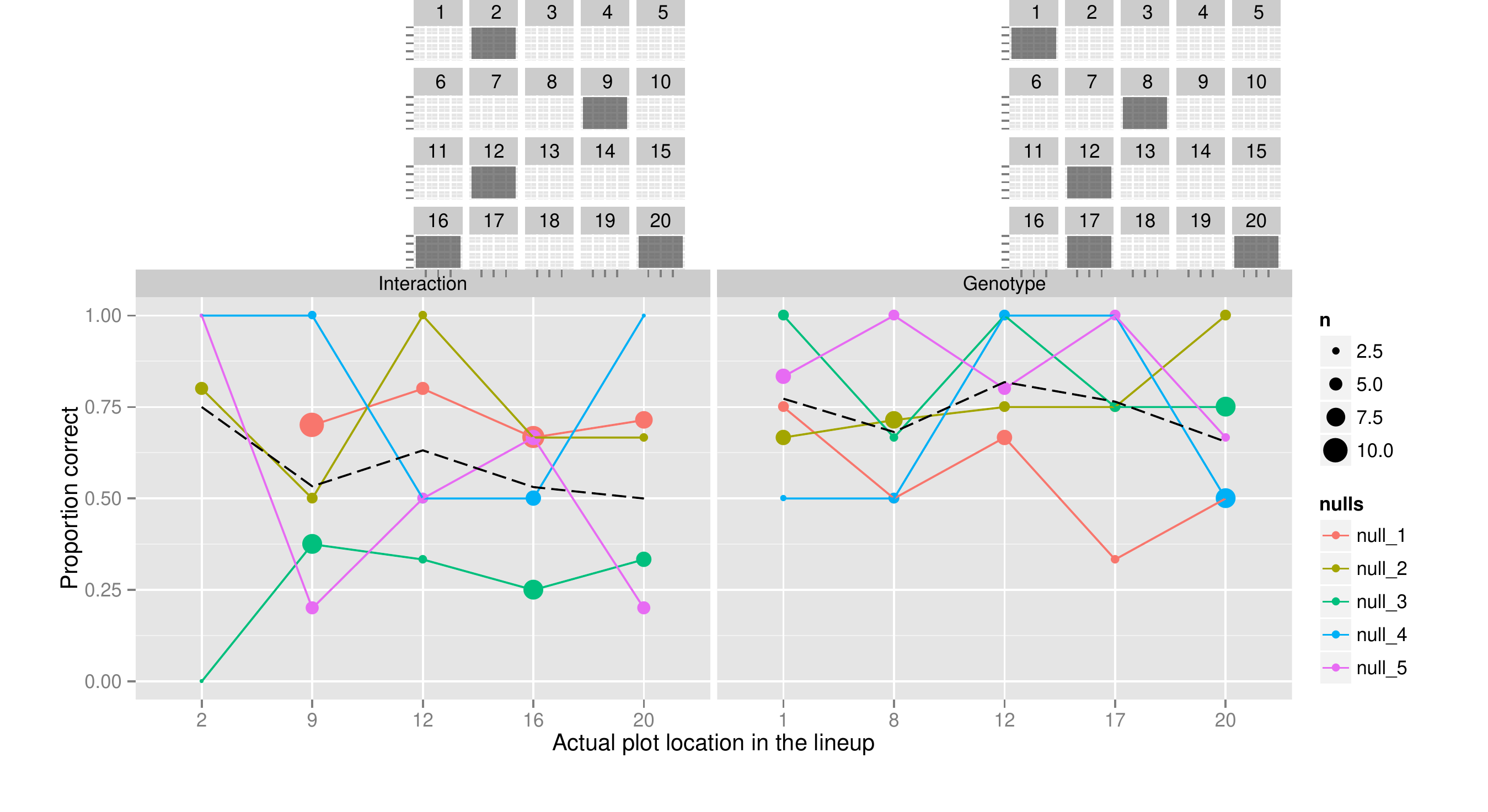} 
   \caption{Location of data plot in the lineup and proportion correct for both Interaction and Genotype effect. Each colored line represents a null set and the size of the dots represents number of responses. The overall average proportions are shown by dashed line. The actual data plot locations are shaded grey on the top panels to demonstrate their relative positions on a lineup.}
   \label{fig:location_effect}
\end{figure}

The detection rate for each data plot location is shown in Figure \ref{fig:location_effect} colored by null sets. The overall average detection rate for each location are shown using dashed lines. Size of the point represents the number of responses. For some locations we have as many as 10 responses. For location 1, we did not have any response for null set 1 in one of the interaction lineups.

We observe some variability of performance for different null sets even though the same actual data plot was used for all these null sets. The lineup protocol is using a finite and small sample (19 in these experiments) of all possible null sets, so it is posisble to see some variability in performance depending on null set. If null sets come from the ``extremes'' of the sampling distribution they may have structure that is more extreme than the actual data plot, making the lineup more difficult to evaluate. From the plot we can see that null set 3 appeared more difficult in the Interaction lineup, because the detection rate was lower regardless of the position of the actual data plot. This null set effect was tested, and results in significant difference only for null set 3 of interaction lineup. The rest of the null sets don't show any differences at 1\% significance level. 


Model \eqref{manova} was fit to the data to test if the detection rates were significantly different for the different locations. For this we use {\it manova()} function of {\it stats} package of \cite{R:2012}. The results are shown in Table \ref{tbl:manova}. The $p$-values for both Interaction and Genotype effect are much bigger than the conventional threshold of 0.05. This suggest that location of the actual data plot in the lineup does not significantly affect performance. 

\begin{table}[hbtp]
\caption{The summary of the results obtained by fitting MANOVA model \eqref{manova} using Wilks test.}
\begin{center}
\begin{tabular}{ccccccccc}
  \hline \hline
 Location & && & \multicolumn{3}{c} {Degrees of Freedom}  & F test \\
 \cline{5-7}
 Effect & DF & Wilks & Approx. F & Numerator& Denumerator &Residual & $p$-value\\
  \hline
  Interaction& 3&0.07986 & 0.592 &15&5.92 & 6 & 0.8082 \\ 
  Genotype & 4&0.01689 & 1.722 & 20&14.22 & 8 & 0.1488 \\ 
   \hline
\end{tabular}
\end{center}
\label{tbl:manova}
\end{table}

We also examined whether the proportion of correct responses differs if the actual plot is on the outer boundary or in the inner locations. The locations 7,8,9,12,13,14 in the lineup are considered inner locations and the rest of the boundary locations are considered to be outer location. As we see in Figure \ref{fig:location_effect}, location 9, 12 are inside for Interaction effect and location 8, 12 are inside for Genotype. It does not show any differences whether the actual plot is inside or outer border of the lineup. We also fitted model \eqref{manova} with two locations, inner and outer, as covariate and observed no significant differences.

%
%

\section{Conclusion}

Human demographics have the potential to influence performance when the lineup protocol is used for statistical inference. In our study of the demographic effects on performance across a set of different experiments we found some statistically significant effects. Age group 36-40, Countries other than India and United states, People who have a graduate degree have a significantly higher detection rate.  Gender does not have any significant effect. However, the effects are minimal, on the order of a few percentage points different from the average. This result is very important for the power of visual test as it demonstrates the robustness of the test for different human factors.


Individual learning trend is observed in time taken but not so much in observer performance. Some individuals improved the performances while others showed decrease in their performances over attempts. This could be interpreted as good, in the sense that it means that observers could realisitically be recruited from source like MTurk, and that substantial training is not necessary in order to obtain useful evaulations of lineups in practice. However, we had hoped that lineups may be a useful teaching tool, to improve students ability to read data plots, and the lack of a learning trend diminishes this idea. 


The simulation experiment reveals that there is no significant effect of location of the actual data plot in the lineup. This is important as the visual statistical inference procedure prescribes that the data plot be placed at random in the lineup. This paper suggests that any random place in a lineup is as good as other places in the lineup. Even though there are variations on the performance depending on different null sets, their impact on probability to correctly evaluate a lineup is very negligible.

There are many more ways that the lineup protocol might be tested to learn what we might expect about its performance for tackling real data mining problems. Possible changes in the lineup protocol in the pipeline are allowing observers to select more than one plot, and to vary the size of the lineup from 20 to a smaller number, and have observers all see different lineups with different null sets.  This paper assessed the effect of human factors on the experiments conducted to date. As changes to the lineup protocol are suggested by other experiments the human factor effects me need to be examined again.

\paragraph{Acknowledgments}
This work was funded in part by National Science Foundation grant DMS 1007697. All studies were conducted with approval from the Institutional Review Board IRB 10-347.

\bibliographystyle{asa}
\bibliography{references}

\end{document}